\documentclass[twocolumn,superscriptaddress,pra,longbibliography,preprintnumbers,amsmath,amssymb,floatfix]{revtex4-1}
\pdfoutput=1

\usepackage{amsmath, amsthm, amssymb, amsfonts}

\usepackage{siunitx}
\usepackage{lmodern}

\usepackage[utf8]{inputenc}	
\usepackage[T1]{fontenc}	
\usepackage{xcolor}		
\usepackage[colorlinks = true,
            linkcolor = purple,
            urlcolor  = blue,
            citecolor = cyan,
            anchorcolor = black]{hyperref}	
\usepackage{microtype}		
\usepackage{float}			
\usepackage{graphicx}

\begin{document}
\title{Single-shot carrier-envelope-phase measurement in ambient air}
\date{\today}	

\author{M.~Kubullek}
\affiliation{Physics Department, Ludwig-Maximilians-Universität Munich, Am Coulombwall 1, 85748 Garching, Germany }

\author{Z.~Wang}
\affiliation{Physics Department, Ludwig-Maximilians-Universität Munich, Am Coulombwall 1, 85748 Garching, Germany }
\affiliation{Max-Planck-Institut für Quantenoptik, Hans-Kopfermann-Straße 1, 85748 Garching, Germany}

\author{K.~von~der~Brelje}
\affiliation{Physics Department, Ludwig-Maximilians-Universität Munich, Am Coulombwall 1, 85748 Garching, Germany }

\author{D.~Zimin}
\affiliation{Physics Department, Ludwig-Maximilians-Universität Munich, Am Coulombwall 1, 85748 Garching, Germany }
\affiliation{Max-Planck-Institut für Quantenoptik, Hans-Kopfermann-Straße 1, 85748 Garching, Germany}

\author{P.~Rosenberger}
\affiliation{Physics Department, Ludwig-Maximilians-Universität Munich, Am Coulombwall 1, 85748 Garching, Germany }

\author{J.~Schötz}
\affiliation{Physics Department, Ludwig-Maximilians-Universität Munich, Am Coulombwall 1, 85748 Garching, Germany }
\affiliation{Max-Planck-Institut für Quantenoptik, Hans-Kopfermann-Straße 1, 85748 Garching, Germany}

\author{M.~Neuhaus}
\affiliation{Physics Department, Ludwig-Maximilians-Universität Munich, Am Coulombwall 1, 85748 Garching, Germany }

\author{S.~Sederberg}
\affiliation{Joint Attosecond Science Laboratory, National Research Council of Canada and University of Ottawa, Ottawa, Ontario K1A0R6, Canada}

\author{A.~Staudte}
\affiliation{Joint Attosecond Science Laboratory, National Research Council of Canada and University of Ottawa, Ottawa, Ontario K1A0R6, Canada}

\author{N.~Karpowicz}
\affiliation{Max-Planck-Institut für Quantenoptik, Hans-Kopfermann-Straße 1, 85748 Garching, Germany}

\author{M.~F.~Kling}
\affiliation{Physics Department, Ludwig-Maximilians-Universität Munich, Am Coulombwall 1, 85748 Garching, Germany }
\affiliation{Max-Planck-Institut für Quantenoptik, Hans-Kopfermann-Straße 1, 85748 Garching, Germany}

\author{B.~Bergues}
\email[Corresponding author: ]{boris.bergues@mpq.mpg.de}
\affiliation{Physics Department, Ludwig-Maximilians-Universität Munich, Am Coulombwall 1, 85748 Garching, Germany }
\affiliation{Max-Planck-Institut für Quantenoptik, Hans-Kopfermann-Straße 1, 85748 Garching, Germany}
\affiliation{Joint Attosecond Science Laboratory, National Research Council of Canada and University of Ottawa, Ottawa, Ontario K1A0R6, Canada}

\begin{abstract}
The ability to measure and control the carrier envelope phase (CEP) of few-cycle laser pulses is of paramount importance for both frequency metrology and attosecond science.  Here, we present a phase meter relying on the CEP-dependent photocurrents induced by circularly polarized few-cycle pulses focused between electrodes in ambient air. The new device facilitates compact single-shot, CEP measurements under ambient conditions and promises CEP tagging at repetition rates orders of magnitude higher than most conventional CEP detection schemes as well as straightforward implementation at longer wavelengths. 
\end{abstract}

\maketitle

Laser sources for near single-cycle pulses in the near-infrared \cite{Nisoli:1997} and infrared \cite{Gu:2009} developed in the past two decades allow the study of light-matter interactions with a temporal resolution reaching a few tens of attoseconds, well below the period of an optical cycle \cite{Baltuska:2003}. One of the keys for achieving such high temporal resolution is the ability to control the carrier-envelope phase (CEP) of the laser pulses.  Mathematically, the electric field of a Fourier-limited laser pulse, propagating in the $z$-direction can be described as: 
\begin{equation*}
E(t) = \frac{E_0 e^{- \frac{t^2}{\tau^2}}}{\sqrt{1 + \varepsilon^2}} \left[ \cos\left( \omega t + \phi \right), \varepsilon \sin\left( \omega t + \phi \right), 0 \right] \;,
\end{equation*}
where $E_0$  is the electric field amplitude,  $\varepsilon$ the ellipticity, $\omega$ the carrier frequency, $\tau$ the pulse duration, and $\phi$  the CEP. In attosecond science, sub-cycle temporal resolution is achieved by the nonlinear gate induced by the strongest cycle in a few cycle pulse. While the pulse envelope remains rather stable from shot to shot, the CEP is prone to vary due to fluctuations of dispersion, caused by changes in path length, and pump energy experienced by consecutive pulses in a pulse train. Fluctuations of the CEP translate into a time jitter of the temporal gate by about half a period of the driving light pulse, thus deteriorating the temporal resolution.

Therefore, it is of importance to measure and stabilize the CEP accordingly. Several schemes have been devised to measure the CEP. The f-2f technique for example relies on the spectral interference of the fundamental and second harmonic of sufficiently broadband fields \cite{Holzwarth:2000,Reichert:1999,Telle:1999,Xu:1996}. While recent progress has facilitated single-shot CEP-measurements at a central wavelength around \SI{800}{\nano\meter} \cite{Lucking:2014}, the f-2f technique is limited by the availability of spectrometers, and thus not easily transposable to different wavelength ranges and limited in acquisition rate (so far to \SI{10}{\kilo\hertz} \cite{Ren:2017}). A variant of the f-2f setup where the grating spectrometer is replaced by temporal dispersion in a \SI{}{\kilo\meter} long fiber and a fast photodiode detection (TOUCAN) removes some of these limitations while relying on a careful selection of the dispersive medium \cite{Kurucz:2019}.

Another well-established, and widely used technique in the last decade relies on above threshold ionizatin (ATI) of a rare gas atoms (typically xenon). While the use of ATI was initially proposed for both circularly \cite{Dietrich:2000} and linearly \cite{Paulus:2001} polarized pulses, its implementation known as the stereo-ATI phase-meter \cite{Paulus:2003} relies on time-of-flight (TOF) measurements of ATI recollision electrons ionized by linearly polarized pulses. This technique has facilitated single-shot CEP-measurement \cite{Wittmann:2009}, at repetition rates up to \SI{100}{\kilo\hertz} \cite{Hoff:2018}, and has allowed major breakthroughs in the study of field-driven dynamics in atoms \cite{Bergues:2012,Schoffler:2016,Bergues:2015,Kubel:2016}, molecules \cite{Alnaser:2014,Miura:2014,Kubel:2016-2}, nanostructures \cite{Sussmann:2011,Sussmann:2015} and solids \cite{Kormin:2018,Kessel:2018}. Despite its great success, the stereo-ATI phase meter is a rather sophisticated apparatus relying on ultra-high vacuum components, and microchannel-plate detection. The main reason for the complexity is the need for an electron-TOF measurement, which is only possible under ultra-high vacuum. Additionally, the unfortunate scaling of the recollision probability with wavelength of $\lambda^{-5}$ to $\lambda^{-6}$ \cite{Tate:2007}, impedes the extension of the stereo-ATI phase meter to longer wavelengths.

Therefore, the development of a more compact, single-shot CEP measurement technique, that can be reduced in terms of complexity, and extended in its wavelength range (and potentially operate at higher repetition rates than the stereo-ATI phase meter), is highly desirable. It has been demonstrated that strong field exitation, and ballistic light field acceleration of the conduction band population in a wide bandgap solid can produce an electric current, whose direction and amplitude relate to the CEP of the laser pulse \cite{Schiffrin:2013}. These currents can be detected using electronic amplifiers enabling the measurement of the CEP\cite{Paasch-Colberg:2014}. Single shot sensitivity has, however, so far not been achieved with this technique. 

The use of circularly polarized input pulses offers several advantages over a linearly polarized input\cite{Dietrich:2000,Bergues:2012-2,Fukahori:2017}. During strong-field ionization of a single atom in a circularly polarized laser pulse, electrons are preferentially emitted in the polarization plane. When the Coulomb interaction with the ionic core is neglected, their drift momentum is perpendicular to the direction of the maximum electric field. Thus, the CEP can be directly retrieved from the preferred electron emission direction, since in this case it coincides with the angle of the maximum electric field. In general, the emission direction will coincide with the CEP up to a constant offset value \cite{Bergues:2012-2}. Most importantly, no time-of-flight measurement is required with a circular-polarization phase-meter (CP-phase-meter), which allows the implementation of much simpler CEP-measurement devices. ATI based CEP-measurements using circularly polarized pulses have been simulated\cite{Bergues:2012-2} and tested experimentally\cite{Debrah:2019}. Alternatively, it has been shown that CEP measurements can also be done by sampling the \si{\tera\hertz} pulses emitted from laser generated ambient air plasma\cite{Kress:2006, Dai:2009,Bai:2012}. Here, we demonstrate a compact single-shot CP-phase-meter relying on the measurement of transient electrical currents in ambient air plasma. This new device is the potentially simplest conceivable implementation of the CP-phase-meter\cite{Bergues:2012-2}. Combining the advantages of circular polarization and electric detection, it enables a straightforward, single-shot CEP measurement under ambient conditions. The acquisition rate is only limited by the bandwidth of the high-gain electric amplifiers (currently MHz rates). The fact that the concept relies on direct ionization makes it easily extendable to pulses with longer wavelengths with comparable peak intensities.


\section*{Experimental Setup}
\begin{figure}[htb]
	\centering
	\includegraphics[width=\columnwidth]{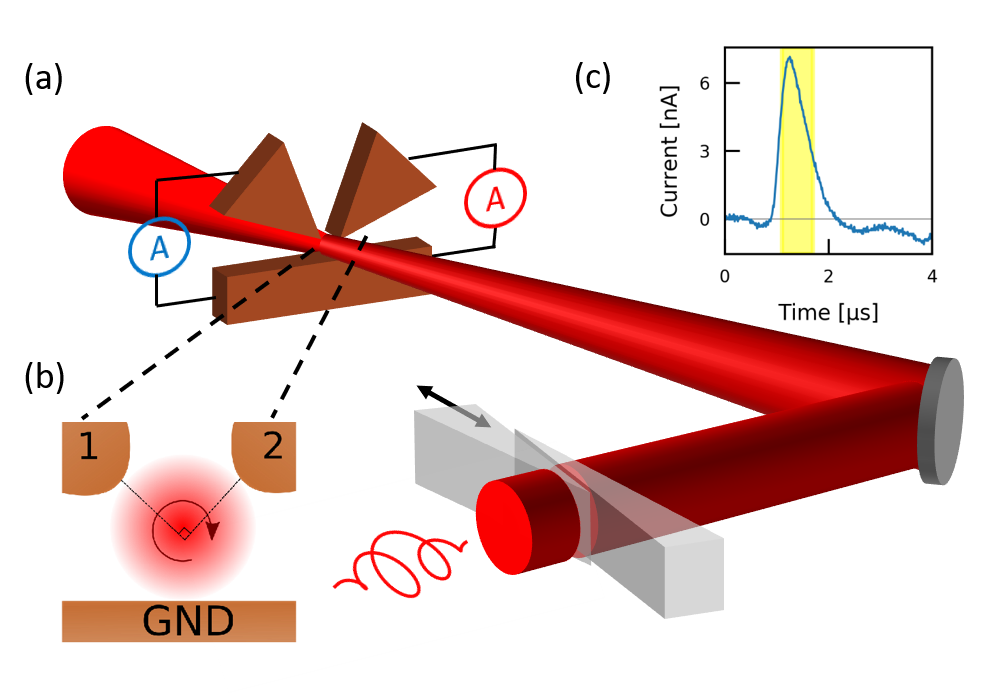}
	\caption{(a)~Experimental setup. Few-cycle laser pulses are sent through a quarter-wave plate to convert their linear polarization to near circular, and focused into ambient air in the gap between three metal electrodes: two tip-shaped electrodes and a ground electrode. A movable pair of wedges is used to change the dispersion in the beam path. (b)~Detailed view of the focus position in the gap between the electrodes. (c)~Single-shot current signal flowing between the tips and the ground. The signal is integrated over the yellow region.}
	\label{fig:ExpSetup}
\end{figure}

The experimental setup for the single-shot CEP characterization in air is shown in Fig.~\hyperref[fig:ExpSetup]{1~(a)}.

Circularly polarized few-cycle laser pulses are focused to a spot size of \SI{32}{\micro\meter} full width at half maximum (FWHM) in between three metal electrodes: two tip-shaped electrodes separated by \SI{60}{\micro\meter} and a third larger, planar electrode positioned \SI{90}{\micro\meter} below the two tips (see Fig.~\hyperref[fig:ExpSetup]{1~(b)}). In the focus, the laser pulses reach peak intensities of about \SI{2e15}{\watt\per\centi\meter\squared} and ionize ambient air, inducing a transient current. For each laser pulse, the CEP-dependent direction of the transient current vector is probed by measuring the currents $I_1$ and $I_2$ flowing between each of the two tips and the ground electrode. The currents are amplified by a factor of \SI{e7}{\volt/\ampere} with a transimpedance amplifier. The two amplified single shot signals (cf. Fig.~\hyperref[fig:ExpSetup]{1~(c)}), one for each tip (blue and red circuits in Fig.~\hyperref[fig:ExpSetup]{1~(a)}), are then integrated using a boxcar integrator. The boxcar DC voltage outputs $Q_1$ and $Q_2$, which are proportional to the charges flowing in the two circuits, are recorded for each laser shot using a DAQ-card.

The laser system used in the present study is a \SI{10}{\kilo\hertz} titanium:sapphire chirped pulse amplification (CPA) system (Spectra Physics Femtopower HR CEP4) that delivers CEP stable (down to ca. \SI{100}{\milli\radian} rms \cite{Lucking:2014}) pulses with \SI{700}{\micro\joule} pulse energy, sub‑\SI{25}{\femto\second} pulse duration and a central wavelength of about \SI{780}{\nano\meter}. The output pulses are spectrally broadened in a gas-filled hollow-core fiber and compressed with a combination of chirped mirrors and fused silica wedges to sub-two cycle duration, typically \SI{4}{\femto\second} (FWHM intensity envelope). The central wavelength is \SI{750}{\nano\meter}. Pulses with an energy of about \SI{100}{\micro\joule} are sent through a broadband quarter-wave plate to convert their polarization from linear to near circular ($\varepsilon = 0.84$) and are focused in between the electrodes with a spherical silver mirror ($f=\SI{350}{\milli\meter}$). The CEP is controlled by changing the dispersion in the stretcher of the CPA multi-pass amplifier.

\section*{Result}
\begin{figure}[htb]
	\centering
	\includegraphics[width=0.9\columnwidth]{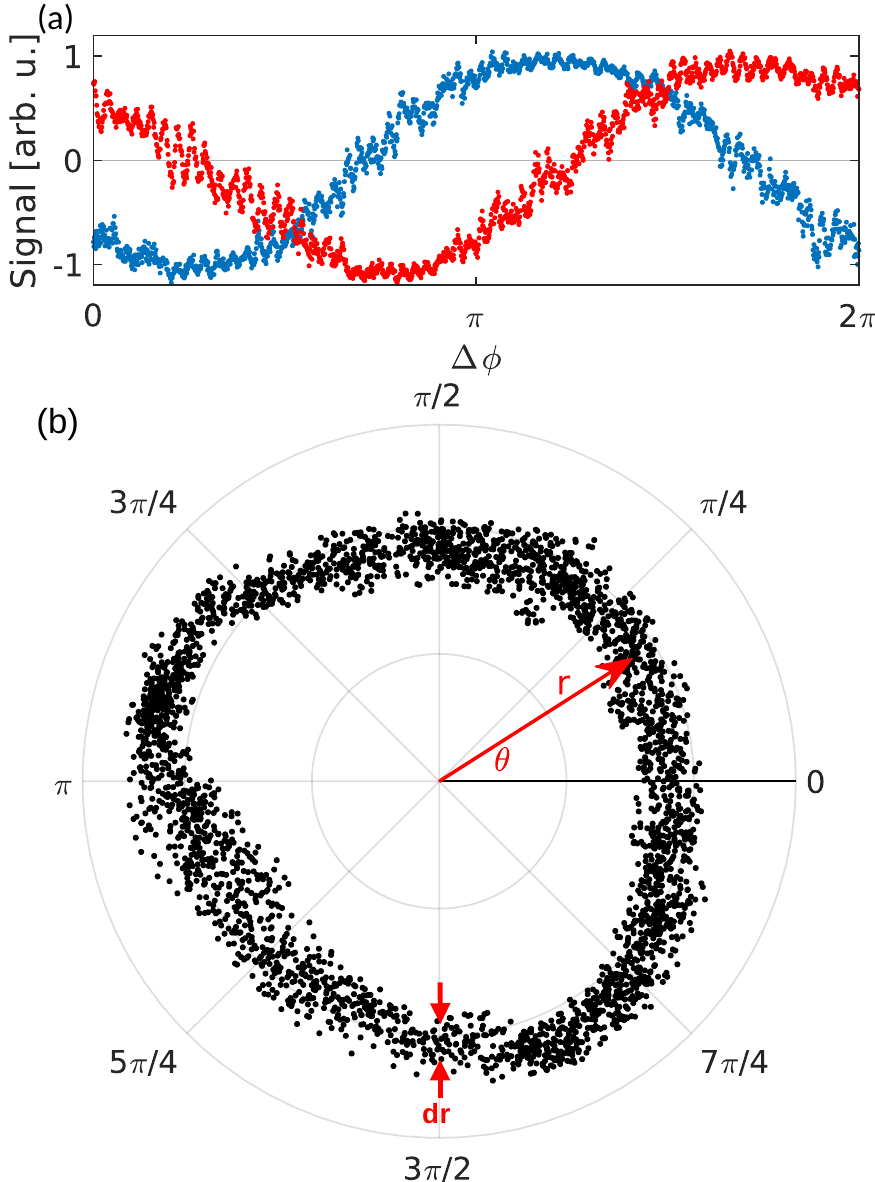}
	\caption{(a)~Single shot signal $Q_1$ (blue) and $Q_2$ (red) recorded while linearly scanning the CEP from $0$ to $2\pi$. (b)~Single-shot parametric plot recorded while scanning the CEP from $0$ to $2\pi$ and back. The standard deviation $dr$ of the radius of the parametric plot is indicated by the two red arrows.}
	\label{fig:Potato}
\end{figure}

The measured signals $Q_1$ and $Q_2$ are plotted in Fig.~\hyperref[fig:Potato]{2~(a)} for a series of \num{1650} consecutive laser shots recorded while linearly changing the CEP from $0$ to $2\pi$. 

Both signals were centered by subtraction of the CEP averaged value and normalized in amplitude. Note that both the offset subtraction and the normalization do not require a stable CEP and can be performed in the same way for a pulse sequence with a randomly fluctuating CEP. The CEP dependent signals $Q_1$ and $Q_2$ oscillate out of phase with a phase shift of \SI{92}{\degree}, close to what is expected for perfect positioning of the laser focus, where the angle between the lines connecting the injection point with the two electrodes spans \SI{90}{\degree} \cite{Bergues:2012-2}. 

As for stereo-ATI phase-meter measurements, $Q_1$ and $Q_2$ can be plotted parametrically as a function of their polar angle $\theta = \arctan \! 2(Q_2, Q_1)$ and $r = \sqrt{Q_1^2 + Q_2^2}$  (see Fig.~\hyperref[fig:Potato]{2~(b)}). The quantity $dr/r= \SI{0.107}{\radian}$ provides a lower limit for the uncertainty of the measurement in Fig. \hyperref[fig:Potato]{2~(b)} \cite{Kubel:2012}. While the CEP is a monotonic function $\phi(\theta)$ of the polar angle $\theta$ in the parametric plot of Fig.~\hyperref[fig:Potato]{2~(b)}, this function is not necessarily linear. Deviation from a linear relation may have different causes, including a slight ellipticity of the input pulse polarization, and a focus that is not perfectly centered in the gap. This is analogous to the stereo-ATI phase meter, where the shape of the parametric plot, depends on the exact experimental conditions such as the position of the TOF integration gates \cite{Kubel:2012}. Fortunately, in either case, the exact shape of the parametric plot is not important for the measurement as the CEP can be retrieved from the polar angle via a rebinning procedure\cite{Wittmann:2009}. The latter relies on the assumption that all CEP values are equally probable within the CEP scan, which is well fulfilled in the present experiment. The dependence of the CEP on the polar angle  is then simply obtained by sorting the polar angles in ascending order over the range of the scan and mapping them onto a linear CEP interval from $-\pi$ to $\pi$. 
\begin{figure}[htb]
	\centering
	\includegraphics[width=0.9\columnwidth]{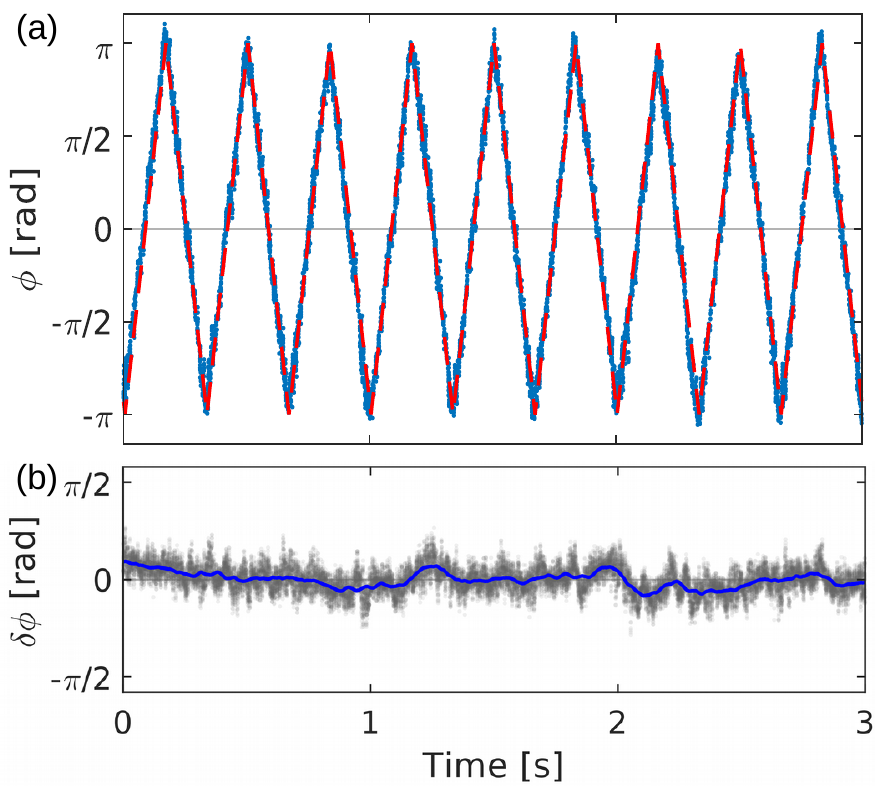}
	\caption{(a)~Measured CEP while sweeping it with a triangular function from $-\pi$ to $\pi$. The dashed red line was obtained by fitting the constant phase of the \SI{3}{\hertz} triangle function to the data points. (b)~Difference between the fitted and measured triangular wave. The blue line represents the averaged value over \SI{100}{\milli\second}}
	\label{fig:RetCEP}
\end{figure}

In order to determine the precision of the measurement, we compare in Fig.~\hyperref[fig:RetCEP]{3~(a)} the retrieved CEP to its nominal value, which (for a perfectly stable CEP) is inferred from the known dispersion introduced in the stretcher. The latter is varied as a triangular function of time to generate a uniform CEP distribution between $-\pi$ and $\pi$. The calibration function $\phi(\theta)$ was determined for each oscillation period of the triangular waveform with the method described above. An upper limit for the uncertainty of the measurement is calculated as the standard deviation of the difference between the measured and the nominal CEP curves (shown in Fig.~\hyperref[fig:RetCEP]{3~(b)}). For the data of Fig.~\hyperref[fig:Potato]{2~(b)} we obtain an upper limit of \SI{206}{\milli\radian}. On longer time scales, the accuracy evolves from \SI{211}{\milli\radian} on the time scale of a few seconds (data of Fig.~\hyperref[fig:RetCEP]{3~(a)}) to \SI{356}{\milli\radian} for an acquisition time of one minute. 

\begin{figure}[htb]
	\centering
	\includegraphics[width=0.9\columnwidth]{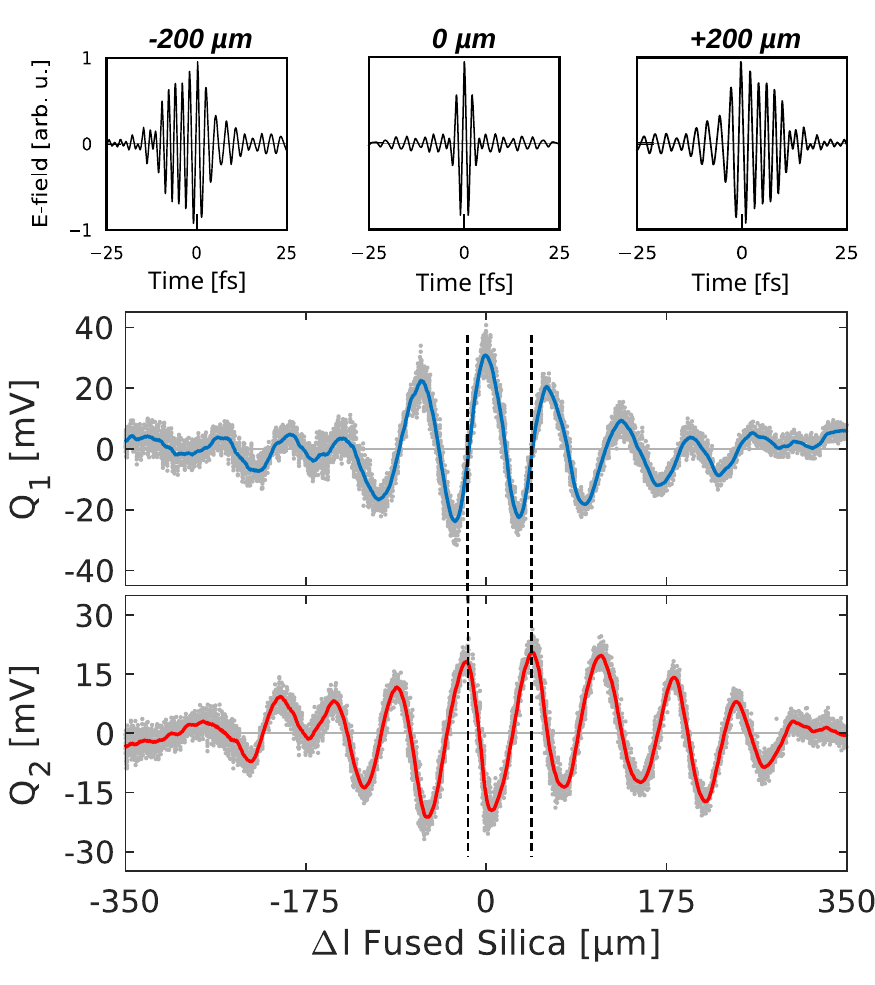}
	\caption{Lower panels: raw single shot signals (gray scatter plots) and averaged signal (solid lines) recorded while linearly changing the amount of dispersive material in the beam path. The signals exhibit a phase shift close to \SI{90}{\degree}, indicated by the dashed lines. Upper panel: Fourier-limited pulse (\SI{3.5}{\femto\second} intensity FWHM) calculated from the measured laser spectrum (center) and simulated pulses after propagation through \SI{200}{\micro\meter} less (left) or \SI{200}{\micro\meter} more (right) fused silica (\SI{9.3}{\femto\second} intensity FWHM).}
	\label{fig:DScan}
\end{figure}

Even though the stability of the measurement and the signal to noise ratio can still be improved, the performance of the new CP-phase-meter is already comparable to that of the stereo-ATI phase-meter. While the f-2f technique only provides information on the CEP, the CP-phase-meter naturally yields information about the pulse duration. Importantly the sensitivity of the measurement increases towards shorter pulse durations, while still supporting measurements with \SI{10}{\femto\second} pulses. This is illustrated in Fig.\ref{fig:DScan}, where the signals $Q_1$ and $Q_2$ are plotted as a function of the pulse propagation distance through glass. 

The most important asset of the new technique, besides its striking simplicity, is its potential for single shot CEP-measurements at much higher repetition rates than achievable with today’s techniques. Unlike the stereo-ATI phase meter, which is intrinsically limited to a few hundred kHz by the time of flight measurement, the new technique, is only limited by the gain-bandwidth product of the amplifier. Given the \SI{2}{\micro\second} duration of the amplified current signal (cf. Fig.~\hyperref[fig:ExpSetup]{1~(c)}), the technique can be readily implemented at more than \SI{100}{\kilo\hertz} with commercially available integrators. We expect that further improvement of the signal to noise ratio by better shielding and tighter focusing will facilitate its implementation at \si{\mega\hertz} repetition rates.

\section*{Summary and Conclusion}
We have demonstrated a simple implementation of the circular polarization phase-meter, which enables single-shot CEP measurement with a precision of about \SI{200}{\milli\radian}. While the performance of our prototype is comparable to that of the widespread stereo-ATI phase-meter, its complexity is dramatically reduced since it only consists of a centimeter-sized setup that works in ambient air.  Since the measurement rate is only limited by the bandwidth of the current amplifier, the technique can easily be applied at a repetition rate of \SI{100}{\kilo\hertz} and beyond. In addition, since the CP-phase meter does not rely on the recollision process, it is also applicable at longer wavelengths. The new technique thus represents an appealing alternative to the rather complex ultra-high vacuum apparatus used nowadays for single-shot CEP detection.  

\subsection*{Funding}
German Research Foundation (KL-1439/11-1, SFB NOA). Max Planck Society (MP Fellow program).

\subsection*{Acknowledgements}
We thank Hartmut Schröder, Matthias Kübel, Aleksey Korobenko, Kyle Johnston, and Valentina Shumakova for fruitful discussions and are grateful for support by David Villeneuve, Paul Corkum and Ferenc Krausz.\\

\footnotesize
\bibliography{references}

\begin{thebibliography}{36}%
\makeatletter
\providecommand \@ifxundefined [1]{%
 \@ifx{#1\undefined}
}%
\providecommand \@ifnum [1]{%
 \ifnum #1\expandafter \@firstoftwo
 \else \expandafter \@secondoftwo
 \fi
}%
\providecommand \@ifx [1]{%
 \ifx #1\expandafter \@firstoftwo
 \else \expandafter \@secondoftwo
 \fi
}%
\providecommand \natexlab [1]{#1}%
\providecommand \enquote  [1]{``#1''}%
\providecommand \bibnamefont  [1]{#1}%
\providecommand \bibfnamefont [1]{#1}%
\providecommand \citenamefont [1]{#1}%
\providecommand \href@noop [0]{\@secondoftwo}%
\providecommand \href [0]{\begingroup \@sanitize@url \@href}%
\providecommand \@href[1]{\@@startlink{#1}\@@href}%
\providecommand \@@href[1]{\endgroup#1\@@endlink}%
\providecommand \@sanitize@url [0]{\catcode `\\12\catcode `\$12\catcode
  `\&12\catcode `\#12\catcode `\^12\catcode `\_12\catcode `\%12\relax}%
\providecommand \@@startlink[1]{}%
\providecommand \@@endlink[0]{}%
\providecommand \url  [0]{\begingroup\@sanitize@url \@url }%
\providecommand \@url [1]{\endgroup\@href {#1}{\urlprefix }}%
\providecommand \urlprefix  [0]{URL }%
\providecommand \Eprint [0]{\href }%
\providecommand \doibase [0]{http://dx.doi.org/}%
\providecommand \selectlanguage [0]{\@gobble}%
\providecommand \bibinfo  [0]{\@secondoftwo}%
\providecommand \bibfield  [0]{\@secondoftwo}%
\providecommand \translation [1]{[#1]}%
\providecommand \BibitemOpen [0]{}%
\providecommand \bibitemStop [0]{}%
\providecommand \bibitemNoStop [0]{.\EOS\space}%
\providecommand \EOS [0]{\spacefactor3000\relax}%
\providecommand \BibitemShut  [1]{\csname bibitem#1\endcsname}%
\let\auto@bib@innerbib\@empty
\bibitem [{\citenamefont {Nisoli}\ \emph {et~al.}(1997)\citenamefont {Nisoli},
  \citenamefont {Silvestri}, \citenamefont {Svelto}, \citenamefont
  {Szip\"{o}cs}, \citenamefont {Ferencz}, \citenamefont {Spielmann},
  \citenamefont {Sartania},\ and\ \citenamefont {Krausz}}]{Nisoli:1997}%
  \BibitemOpen
  \bibfield  {author} {\bibinfo {author} {\bibfnamefont {M.}~\bibnamefont
  {Nisoli}}, \bibinfo {author} {\bibfnamefont {S.~De}\ \bibnamefont
  {Silvestri}}, \bibinfo {author} {\bibfnamefont {O.}~\bibnamefont {Svelto}},
  \bibinfo {author} {\bibfnamefont {R.}~\bibnamefont {Szip\"{o}cs}}, \bibinfo
  {author} {\bibfnamefont {K.}~\bibnamefont {Ferencz}}, \bibinfo {author}
  {\bibfnamefont {Ch.}\ \bibnamefont {Spielmann}}, \bibinfo {author}
  {\bibfnamefont {S.}~\bibnamefont {Sartania}}, \ and\ \bibinfo {author}
  {\bibfnamefont {F.}~\bibnamefont {Krausz}},\ }\bibfield  {title} {\enquote
  {\bibinfo {title} {Compression of high-energy laser pulses below 5 fs},}\
  }\href {\doibase 10.1364/OL.22.000522} {\bibfield  {journal} {\bibinfo
  {journal} {Optics Letters}\ }\textbf {\bibinfo {volume} {22}},\ \bibinfo
  {pages} {522--524} (\bibinfo {year} {1997})}\BibitemShut {NoStop}%
\bibitem [{\citenamefont {Gu}\ \emph {et~al.}(2009)\citenamefont {Gu},
  \citenamefont {Marcus}, \citenamefont {Deng}, \citenamefont {Metzger},
  \citenamefont {Teisset}, \citenamefont {Ishii}, \citenamefont {Fuji},
  \citenamefont {Baltuska}, \citenamefont {Butkus}, \citenamefont {Pervak},
  \citenamefont {Ishizuki}, \citenamefont {Taira}, \citenamefont {Kobayashi},
  \citenamefont {Kienberger},\ and\ \citenamefont {Krausz}}]{Gu:2009}%
  \BibitemOpen
  \bibfield  {author} {\bibinfo {author} {\bibfnamefont {Xun}\ \bibnamefont
  {Gu}}, \bibinfo {author} {\bibfnamefont {Gilad}\ \bibnamefont {Marcus}},
  \bibinfo {author} {\bibfnamefont {Yunpei}\ \bibnamefont {Deng}}, \bibinfo
  {author} {\bibfnamefont {Thomas}\ \bibnamefont {Metzger}}, \bibinfo {author}
  {\bibfnamefont {Catherine}\ \bibnamefont {Teisset}}, \bibinfo {author}
  {\bibfnamefont {Nobuhisa}\ \bibnamefont {Ishii}}, \bibinfo {author}
  {\bibfnamefont {Takao}\ \bibnamefont {Fuji}}, \bibinfo {author}
  {\bibfnamefont {Andrius}\ \bibnamefont {Baltuska}}, \bibinfo {author}
  {\bibfnamefont {Rytis}\ \bibnamefont {Butkus}}, \bibinfo {author}
  {\bibfnamefont {Volodymyr}\ \bibnamefont {Pervak}}, \bibinfo {author}
  {\bibfnamefont {Hideki}\ \bibnamefont {Ishizuki}}, \bibinfo {author}
  {\bibfnamefont {Takunori}\ \bibnamefont {Taira}}, \bibinfo {author}
  {\bibfnamefont {Takayoshi}\ \bibnamefont {Kobayashi}}, \bibinfo {author}
  {\bibfnamefont {Reinhard}\ \bibnamefont {Kienberger}}, \ and\ \bibinfo
  {author} {\bibfnamefont {Ferenc}\ \bibnamefont {Krausz}},\ }\bibfield
  {title} {\enquote {\bibinfo {title} {Generation of
  carrier-envelope-phase-stable 2-cycle 740-$\mu$j pulses at 2.1-$\mu$m carrier
  wavelength},}\ }\href {\doibase 10.1364/OE.17.000062} {\bibfield  {journal}
  {\bibinfo  {journal} {Optics Express}\ }\textbf {\bibinfo {volume} {17}},\
  \bibinfo {pages} {62--69} (\bibinfo {year} {2009})}\BibitemShut {NoStop}%
\bibitem [{\citenamefont {Baltu{\v s}ka}\ \emph {et~al.}(2003)\citenamefont
  {Baltu{\v s}ka}, \citenamefont {Udem}, \citenamefont {Uiberacker},
  \citenamefont {Hentschel}, \citenamefont {Goulielmakis}, \citenamefont
  {Gohle}, \citenamefont {Holzwarth}, \citenamefont {Yakovlev}, \citenamefont
  {Scrinzi}, \citenamefont {H{\"a}nsch},\ and\ \citenamefont
  {Krausz}}]{Baltuska:2003}%
  \BibitemOpen
  \bibfield  {author} {\bibinfo {author} {\bibfnamefont {A.}~\bibnamefont
  {Baltu{\v s}ka}}, \bibinfo {author} {\bibfnamefont {Th.}\ \bibnamefont
  {Udem}}, \bibinfo {author} {\bibfnamefont {M.}~\bibnamefont {Uiberacker}},
  \bibinfo {author} {\bibfnamefont {M.}~\bibnamefont {Hentschel}}, \bibinfo
  {author} {\bibfnamefont {E.}~\bibnamefont {Goulielmakis}}, \bibinfo {author}
  {\bibfnamefont {Ch.}\ \bibnamefont {Gohle}}, \bibinfo {author} {\bibfnamefont
  {R.}~\bibnamefont {Holzwarth}}, \bibinfo {author} {\bibfnamefont {V.~S.}\
  \bibnamefont {Yakovlev}}, \bibinfo {author} {\bibfnamefont {A.}~\bibnamefont
  {Scrinzi}}, \bibinfo {author} {\bibfnamefont {T.~W.}\ \bibnamefont
  {H{\"a}nsch}}, \ and\ \bibinfo {author} {\bibfnamefont {F.}~\bibnamefont
  {Krausz}},\ }\bibfield  {title} {\enquote {\bibinfo {title} {Attosecond
  control of electronic processes by intense light fields},}\ }\href {\doibase
  10.1038/nature01414} {\bibfield  {journal} {\bibinfo  {journal} {Nature}\
  }\textbf {\bibinfo {volume} {421}},\ \bibinfo {pages} {611--615} (\bibinfo
  {year} {2003})}\BibitemShut {NoStop}%
\bibitem [{\citenamefont {Holzwarth}\ \emph {et~al.}(2000)\citenamefont
  {Holzwarth}, \citenamefont {Udem}, \citenamefont {H\"ansch}, \citenamefont
  {Knight}, \citenamefont {Wadsworth},\ and\ \citenamefont
  {Russell}}]{Holzwarth:2000}%
  \BibitemOpen
  \bibfield  {author} {\bibinfo {author} {\bibfnamefont {R.}~\bibnamefont
  {Holzwarth}}, \bibinfo {author} {\bibfnamefont {Th.}\ \bibnamefont {Udem}},
  \bibinfo {author} {\bibfnamefont {T.~W.}\ \bibnamefont {H\"ansch}}, \bibinfo
  {author} {\bibfnamefont {J.~C.}\ \bibnamefont {Knight}}, \bibinfo {author}
  {\bibfnamefont {W.~J.}\ \bibnamefont {Wadsworth}}, \ and\ \bibinfo {author}
  {\bibfnamefont {P.~St.~J.}\ \bibnamefont {Russell}},\ }\bibfield  {title}
  {\enquote {\bibinfo {title} {Optical frequency synthesizer for precision
  spectroscopy},}\ }\href {\doibase 10.1103/PhysRevLett.85.2264} {\bibfield
  {journal} {\bibinfo  {journal} {Physical Review Letters}\ }\textbf {\bibinfo
  {volume} {85}},\ \bibinfo {pages} {2264--2267} (\bibinfo {year}
  {2000})}\BibitemShut {NoStop}%
\bibitem [{\citenamefont {Reichert}\ \emph {et~al.}(1999)\citenamefont
  {Reichert}, \citenamefont {Holzwarth}, \citenamefont {Udem},\ and\
  \citenamefont {H{\"a}nsch}}]{Reichert:1999}%
  \BibitemOpen
  \bibfield  {author} {\bibinfo {author} {\bibfnamefont {J.}~\bibnamefont
  {Reichert}}, \bibinfo {author} {\bibfnamefont {R.}~\bibnamefont {Holzwarth}},
  \bibinfo {author} {\bibfnamefont {T.}~\bibnamefont {Udem}}, \ and\ \bibinfo
  {author} {\bibfnamefont {T.~W.}\ \bibnamefont {H{\"a}nsch}},\ }\bibfield
  {title} {\enquote {\bibinfo {title} {Measuring the frequency of light with
  mode-locked lasers.}}\ }\href {\doibase 10.1016/S0030-4018(99)00491-5}
  {\bibfield  {journal} {\bibinfo  {journal} {Optics Communications}\ }\textbf
  {\bibinfo {volume} {172}},\ \bibinfo {pages} {59--68} (\bibinfo {year}
  {1999})}\BibitemShut {NoStop}%
\bibitem [{\citenamefont {Telle}\ \emph {et~al.}(1999)\citenamefont {Telle},
  \citenamefont {Steinmeyer}, \citenamefont {Dunlop}, \citenamefont {Stenger},
  \citenamefont {Sutter},\ and\ \citenamefont {Keller}}]{Telle:1999}%
  \BibitemOpen
  \bibfield  {author} {\bibinfo {author} {\bibfnamefont {H.~R.}\ \bibnamefont
  {Telle}}, \bibinfo {author} {\bibfnamefont {G.}~\bibnamefont {Steinmeyer}},
  \bibinfo {author} {\bibfnamefont {A.~E.}\ \bibnamefont {Dunlop}}, \bibinfo
  {author} {\bibfnamefont {J.}~\bibnamefont {Stenger}}, \bibinfo {author}
  {\bibfnamefont {D.H.}\ \bibnamefont {Sutter}}, \ and\ \bibinfo {author}
  {\bibfnamefont {U.}~\bibnamefont {Keller}},\ }\bibfield  {title} {\enquote
  {\bibinfo {title} {Carrier-envelope offset phase control: A novel concept for
  absolute optical frequency measurement and ultrashort pulse generation},}\
  }\href {\doibase 10.1007/s003400050813} {\bibfield  {journal} {\bibinfo
  {journal} {Applied Physics B}\ }\textbf {\bibinfo {volume} {69}},\ \bibinfo
  {pages} {327--332} (\bibinfo {year} {1999})}\BibitemShut {NoStop}%
\bibitem [{\citenamefont {Xu}\ \emph {et~al.}(1996)\citenamefont {Xu},
  \citenamefont {Spielmann}, \citenamefont {Poppe}, \citenamefont {Brabec},
  \citenamefont {Krausz},\ and\ \citenamefont {H\"{a}nsch}}]{Xu:1996}%
  \BibitemOpen
  \bibfield  {author} {\bibinfo {author} {\bibfnamefont {L.}~\bibnamefont
  {Xu}}, \bibinfo {author} {\bibfnamefont {Ch.}\ \bibnamefont {Spielmann}},
  \bibinfo {author} {\bibfnamefont {A.}~\bibnamefont {Poppe}}, \bibinfo
  {author} {\bibfnamefont {T.}~\bibnamefont {Brabec}}, \bibinfo {author}
  {\bibfnamefont {F.}~\bibnamefont {Krausz}}, \ and\ \bibinfo {author}
  {\bibfnamefont {T.~W.}\ \bibnamefont {H\"{a}nsch}},\ }\bibfield  {title}
  {\enquote {\bibinfo {title} {Route to phase control of ultrashort light
  pulses},}\ }\href {\doibase 10.1364/OL.21.002008} {\bibfield  {journal}
  {\bibinfo  {journal} {Optics Letters}\ }\textbf {\bibinfo {volume} {21}},\
  \bibinfo {pages} {2008--2010} (\bibinfo {year} {1996})}\BibitemShut {NoStop}%
\bibitem [{\citenamefont {L\"{u}cking}\ \emph {et~al.}(2014)\citenamefont
  {L\"{u}cking}, \citenamefont {Crozatier}, \citenamefont {Forget},
  \citenamefont {Assion},\ and\ \citenamefont {Krausz}}]{Lucking:2014}%
  \BibitemOpen
  \bibfield  {author} {\bibinfo {author} {\bibfnamefont {Fabian}\ \bibnamefont
  {L\"{u}cking}}, \bibinfo {author} {\bibfnamefont {Vincent}\ \bibnamefont
  {Crozatier}}, \bibinfo {author} {\bibfnamefont {Nicolas}\ \bibnamefont
  {Forget}}, \bibinfo {author} {\bibfnamefont {Andreas}\ \bibnamefont
  {Assion}}, \ and\ \bibinfo {author} {\bibfnamefont {Ferenc}\ \bibnamefont
  {Krausz}},\ }\bibfield  {title} {\enquote {\bibinfo {title} {Approaching the
  limits of carrier-envelope phase stability in a millijoule-class
  amplifier},}\ }\href {\doibase 10.1364/OL.39.003884} {\bibfield  {journal}
  {\bibinfo  {journal} {Optics Letters}\ }\textbf {\bibinfo {volume} {39}},\
  \bibinfo {pages} {3884--3887} (\bibinfo {year} {2014})}\BibitemShut {NoStop}%
\bibitem [{\citenamefont {Ren}\ \emph {et~al.}(2017)\citenamefont {Ren},
  \citenamefont {Summers}, \citenamefont {Pandiri}, \citenamefont {Vajdi},
  \citenamefont {Makhija}, \citenamefont {Fehrenbach}, \citenamefont {Kling},
  \citenamefont {Betsch}, \citenamefont {Wang}, \citenamefont {Kling},
  \citenamefont {Carnes}, \citenamefont {Ben-Itzhak}, \citenamefont
  {Trallero-Herrero},\ and\ \citenamefont {Kumarappan}}]{Ren:2017}%
  \BibitemOpen
  \bibfield  {author} {\bibinfo {author} {\bibfnamefont {Xiaoming}\
  \bibnamefont {Ren}}, \bibinfo {author} {\bibfnamefont {A.~M.}\ \bibnamefont
  {Summers}}, \bibinfo {author} {\bibfnamefont {Kanaka~Raju}\ \bibnamefont
  {Pandiri}}, \bibinfo {author} {\bibfnamefont {Aram}\ \bibnamefont {Vajdi}},
  \bibinfo {author} {\bibfnamefont {Varun}\ \bibnamefont {Makhija}}, \bibinfo
  {author} {\bibfnamefont {C.~W.}\ \bibnamefont {Fehrenbach}}, \bibinfo
  {author} {\bibfnamefont {Nora~G.}\ \bibnamefont {Kling}}, \bibinfo {author}
  {\bibfnamefont {K.~J.}\ \bibnamefont {Betsch}}, \bibinfo {author}
  {\bibfnamefont {Z.}~\bibnamefont {Wang}}, \bibinfo {author} {\bibfnamefont
  {M.~F.}\ \bibnamefont {Kling}}, \bibinfo {author} {\bibfnamefont {K.~D.}\
  \bibnamefont {Carnes}}, \bibinfo {author} {\bibfnamefont {I.}~\bibnamefont
  {Ben-Itzhak}}, \bibinfo {author} {\bibfnamefont {Carlos}\ \bibnamefont
  {Trallero-Herrero}}, \ and\ \bibinfo {author} {\bibfnamefont {Vinod}\
  \bibnamefont {Kumarappan}},\ }\bibfield  {title} {\enquote {\bibinfo {title}
  {Single-shot carrier-envelope-phase tagging using an f{\textendash}2f
  interferometer and a phase meter: a comparison},}\ }\href {\doibase
  10.1088/2040-8986/aa9865} {\bibfield  {journal} {\bibinfo  {journal} {Journal
  of Optics}\ }\textbf {\bibinfo {volume} {19}},\ \bibinfo {pages} {124017}
  (\bibinfo {year} {2017})}\BibitemShut {NoStop}%
\bibitem [{\citenamefont {Kurucz}\ \emph {et~al.}(2019)\citenamefont {Kurucz},
  \citenamefont {T\'{o}th}, \citenamefont {Flender}, \citenamefont {Haizer},
  \citenamefont {Kiss}, \citenamefont {Persielle},\ and\ \citenamefont
  {Cormier}}]{Kurucz:2019}%
  \BibitemOpen
  \bibfield  {author} {\bibinfo {author} {\bibfnamefont {M\'{a}t\'{e}}\
  \bibnamefont {Kurucz}}, \bibinfo {author} {\bibfnamefont {Szabolcs}\
  \bibnamefont {T\'{o}th}}, \bibinfo {author} {\bibfnamefont {Roland}\
  \bibnamefont {Flender}}, \bibinfo {author} {\bibfnamefont {Ludov\'{i}t}\
  \bibnamefont {Haizer}}, \bibinfo {author} {\bibfnamefont {B\'{a}lint}\
  \bibnamefont {Kiss}}, \bibinfo {author} {\bibfnamefont {Benjamin}\
  \bibnamefont {Persielle}}, \ and\ \bibinfo {author} {\bibfnamefont {Eric}\
  \bibnamefont {Cormier}},\ }\bibfield  {title} {\enquote {\bibinfo {title}
  {Single-shot cep drift measurement at arbitrary repetition rate based on
  dispersive fourier transform},}\ }\href {\doibase 10.1364/OE.27.013387}
  {\bibfield  {journal} {\bibinfo  {journal} {Optics Express}\ }\textbf
  {\bibinfo {volume} {27}},\ \bibinfo {pages} {13387--13399} (\bibinfo {year}
  {2019})}\BibitemShut {NoStop}%
\bibitem [{\citenamefont {Dietrich}\ \emph {et~al.}(2000)\citenamefont
  {Dietrich}, \citenamefont {Krausz},\ and\ \citenamefont
  {Corkum}}]{Dietrich:2000}%
  \BibitemOpen
  \bibfield  {author} {\bibinfo {author} {\bibfnamefont {P.}~\bibnamefont
  {Dietrich}}, \bibinfo {author} {\bibfnamefont {F.}~\bibnamefont {Krausz}}, \
  and\ \bibinfo {author} {\bibfnamefont {P.~B.}\ \bibnamefont {Corkum}},\
  }\bibfield  {title} {\enquote {\bibinfo {title} {Determining the absolute
  carrier phase of a few-cycle laser pulse},}\ }\href {\doibase
  10.1364/OL.25.000016} {\bibfield  {journal} {\bibinfo  {journal} {Optics
  Letters}\ }\textbf {\bibinfo {volume} {25}},\ \bibinfo {pages} {16--18}
  (\bibinfo {year} {2000})}\BibitemShut {NoStop}%
\bibitem [{\citenamefont {Paulus}\ \emph {et~al.}(2001)\citenamefont {Paulus},
  \citenamefont {Grasbon}, \citenamefont {Walther}, \citenamefont {Villoresi},
  \citenamefont {Nisoli}, \citenamefont {Stagira}, \citenamefont {Priori},\
  and\ \citenamefont {De~Silvestri}}]{Paulus:2001}%
  \BibitemOpen
  \bibfield  {author} {\bibinfo {author} {\bibfnamefont {G.~G.}\ \bibnamefont
  {Paulus}}, \bibinfo {author} {\bibfnamefont {F.}~\bibnamefont {Grasbon}},
  \bibinfo {author} {\bibfnamefont {H.}~\bibnamefont {Walther}}, \bibinfo
  {author} {\bibfnamefont {P.}~\bibnamefont {Villoresi}}, \bibinfo {author}
  {\bibfnamefont {M.}~\bibnamefont {Nisoli}}, \bibinfo {author} {\bibfnamefont
  {S.}~\bibnamefont {Stagira}}, \bibinfo {author} {\bibfnamefont
  {E.}~\bibnamefont {Priori}}, \ and\ \bibinfo {author} {\bibfnamefont
  {S.}~\bibnamefont {De~Silvestri}},\ }\bibfield  {title} {\enquote {\bibinfo
  {title} {Absolute-phase phenomena in photoionization with few-cycle laser
  pulses},}\ }\href {\doibase 10.1038/35102520} {\bibfield  {journal} {\bibinfo
   {journal} {Nature}\ }\textbf {\bibinfo {volume} {414}},\ \bibinfo {pages}
  {182--184} (\bibinfo {year} {2001})}\BibitemShut {NoStop}%
\bibitem [{\citenamefont {Paulus}\ \emph {et~al.}(2003)\citenamefont {Paulus},
  \citenamefont {Lindner}, \citenamefont {Walther}, \citenamefont
  {Baltu\ifmmode~\check{s}\else \v{s}\fi{}ka}, \citenamefont {Goulielmakis},
  \citenamefont {Lezius},\ and\ \citenamefont {Krausz}}]{Paulus:2003}%
  \BibitemOpen
  \bibfield  {author} {\bibinfo {author} {\bibfnamefont {G.~G.}\ \bibnamefont
  {Paulus}}, \bibinfo {author} {\bibfnamefont {F.}~\bibnamefont {Lindner}},
  \bibinfo {author} {\bibfnamefont {H.}~\bibnamefont {Walther}}, \bibinfo
  {author} {\bibfnamefont {A.}~\bibnamefont {Baltu\ifmmode~\check{s}\else
  \v{s}\fi{}ka}}, \bibinfo {author} {\bibfnamefont {E.}~\bibnamefont
  {Goulielmakis}}, \bibinfo {author} {\bibfnamefont {M.}~\bibnamefont
  {Lezius}}, \ and\ \bibinfo {author} {\bibfnamefont {F.}~\bibnamefont
  {Krausz}},\ }\bibfield  {title} {\enquote {\bibinfo {title} {Measurement of
  the phase of few-cycle laser pulses},}\ }\href {\doibase
  10.1103/PhysRevLett.91.253004} {\bibfield  {journal} {\bibinfo  {journal}
  {Physical Review Letters}\ }\textbf {\bibinfo {volume} {91}},\ \bibinfo
  {pages} {253004} (\bibinfo {year} {2003})}\BibitemShut {NoStop}%
\bibitem [{\citenamefont {Wittmann}\ \emph {et~al.}(2009)\citenamefont
  {Wittmann}, \citenamefont {Horvath}, \citenamefont {Helml}, \citenamefont
  {Sch{\"a}tzel}, \citenamefont {Gu}, \citenamefont {Cavalieri}, \citenamefont
  {Paulus},\ and\ \citenamefont {Kienberger}}]{Wittmann:2009}%
  \BibitemOpen
  \bibfield  {author} {\bibinfo {author} {\bibfnamefont {T.}~\bibnamefont
  {Wittmann}}, \bibinfo {author} {\bibfnamefont {B.}~\bibnamefont {Horvath}},
  \bibinfo {author} {\bibfnamefont {W.}~\bibnamefont {Helml}}, \bibinfo
  {author} {\bibfnamefont {M.~G.}\ \bibnamefont {Sch{\"a}tzel}}, \bibinfo
  {author} {\bibfnamefont {X.}~\bibnamefont {Gu}}, \bibinfo {author}
  {\bibfnamefont {A.~L.}\ \bibnamefont {Cavalieri}}, \bibinfo {author}
  {\bibfnamefont {G.~G.}\ \bibnamefont {Paulus}}, \ and\ \bibinfo {author}
  {\bibfnamefont {R.}~\bibnamefont {Kienberger}},\ }\bibfield  {title}
  {\enquote {\bibinfo {title} {Single-shot carrier-envelope phase measurement
  of few-cycle laser pulses},}\ }\href {https://doi.org/10.1038/nphys1250}
  {\bibfield  {journal} {\bibinfo  {journal} {Nature Physics}\ }\textbf
  {\bibinfo {volume} {5}},\ \bibinfo {pages} {357--362} (\bibinfo {year}
  {2009})}\BibitemShut {NoStop}%
\bibitem [{\citenamefont {Hoff}\ \emph {et~al.}(2018)\citenamefont {Hoff},
  \citenamefont {Furch}, \citenamefont {Witting}, \citenamefont {R\"{u}hle},
  \citenamefont {Adolph}, \citenamefont {Sayler}, \citenamefont {Vrakking},
  \citenamefont {Paulus},\ and\ \citenamefont {Schulz}}]{Hoff:2018}%
  \BibitemOpen
  \bibfield  {author} {\bibinfo {author} {\bibfnamefont {Dominik}\ \bibnamefont
  {Hoff}}, \bibinfo {author} {\bibfnamefont {Federico~J.}\ \bibnamefont
  {Furch}}, \bibinfo {author} {\bibfnamefont {Tobias}\ \bibnamefont {Witting}},
  \bibinfo {author} {\bibfnamefont {Klaus}\ \bibnamefont {R\"{u}hle}}, \bibinfo
  {author} {\bibfnamefont {Daniel}\ \bibnamefont {Adolph}}, \bibinfo {author}
  {\bibfnamefont {A.~Max}\ \bibnamefont {Sayler}}, \bibinfo {author}
  {\bibfnamefont {Marc J.~J.}\ \bibnamefont {Vrakking}}, \bibinfo {author}
  {\bibfnamefont {Gerhard~G.}\ \bibnamefont {Paulus}}, \ and\ \bibinfo {author}
  {\bibfnamefont {Claus~Peter}\ \bibnamefont {Schulz}},\ }\bibfield  {title}
  {\enquote {\bibinfo {title} {Continuous every-single-shot carrier-envelope
  phase measurement and control at 100 khz},}\ }\href {\doibase
  10.1364/OL.43.003850} {\bibfield  {journal} {\bibinfo  {journal} {Optics
  Letters}\ }\textbf {\bibinfo {volume} {43}},\ \bibinfo {pages} {3850--3853}
  (\bibinfo {year} {2018})}\BibitemShut {NoStop}%
\bibitem [{\citenamefont {Bergues}\ \emph {et~al.}(2012)\citenamefont
  {Bergues}, \citenamefont {K{\"u}bel}, \citenamefont {Johnson}, \citenamefont
  {Fischer}, \citenamefont {Camus}, \citenamefont {Betsch}, \citenamefont
  {Herrwerth}, \citenamefont {Senftleben}, \citenamefont {Sayler},
  \citenamefont {Rathje}, \citenamefont {Pfeifer}, \citenamefont {Ben-Itzhak},
  \citenamefont {Jones}, \citenamefont {Paulus}, \citenamefont {Krausz},
  \citenamefont {Moshammer}, \citenamefont {Ullrich},\ and\ \citenamefont
  {Kling}}]{Bergues:2012}%
  \BibitemOpen
  \bibfield  {author} {\bibinfo {author} {\bibfnamefont {Boris}\ \bibnamefont
  {Bergues}}, \bibinfo {author} {\bibfnamefont {Matthias}\ \bibnamefont
  {K{\"u}bel}}, \bibinfo {author} {\bibfnamefont {Nora~G.}\ \bibnamefont
  {Johnson}}, \bibinfo {author} {\bibfnamefont {Bettina}\ \bibnamefont
  {Fischer}}, \bibinfo {author} {\bibfnamefont {Nicolas}\ \bibnamefont
  {Camus}}, \bibinfo {author} {\bibfnamefont {Kelsie~J.}\ \bibnamefont
  {Betsch}}, \bibinfo {author} {\bibfnamefont {Oliver}\ \bibnamefont
  {Herrwerth}}, \bibinfo {author} {\bibfnamefont {Arne}\ \bibnamefont
  {Senftleben}}, \bibinfo {author} {\bibfnamefont {A.~Max}\ \bibnamefont
  {Sayler}}, \bibinfo {author} {\bibfnamefont {Tim}\ \bibnamefont {Rathje}},
  \bibinfo {author} {\bibfnamefont {Thomas}\ \bibnamefont {Pfeifer}}, \bibinfo
  {author} {\bibfnamefont {Itzik}\ \bibnamefont {Ben-Itzhak}}, \bibinfo
  {author} {\bibfnamefont {Robert~R.}\ \bibnamefont {Jones}}, \bibinfo {author}
  {\bibfnamefont {Gerhard~G.}\ \bibnamefont {Paulus}}, \bibinfo {author}
  {\bibfnamefont {Ferenc}\ \bibnamefont {Krausz}}, \bibinfo {author}
  {\bibfnamefont {Robert}\ \bibnamefont {Moshammer}}, \bibinfo {author}
  {\bibfnamefont {Joachim}\ \bibnamefont {Ullrich}}, \ and\ \bibinfo {author}
  {\bibfnamefont {Matthias~F.}\ \bibnamefont {Kling}},\ }\bibfield  {title}
  {\enquote {\bibinfo {title} {Attosecond tracing of correlated
  electron-emission in non-sequential double ionization},}\ }\href
  {https://doi.org/10.1038/ncomms1807} {\bibfield  {journal} {\bibinfo
  {journal} {Nature Communications}\ }\textbf {\bibinfo {volume} {3}},\
  \bibinfo {pages} {813--818} (\bibinfo {year} {2012})}\BibitemShut {NoStop}%
\bibitem [{\citenamefont {Sch\"offler}\ \emph {et~al.}(2016)\citenamefont
  {Sch\"offler}, \citenamefont {Xie}, \citenamefont {Wustelt}, \citenamefont
  {M\"oller}, \citenamefont {Roither}, \citenamefont {Kartashov}, \citenamefont
  {Sayler}, \citenamefont {Baltuska}, \citenamefont {Paulus},\ and\
  \citenamefont {Kitzler}}]{Schoffler:2016}%
  \BibitemOpen
  \bibfield  {author} {\bibinfo {author} {\bibfnamefont {Markus~S.}\
  \bibnamefont {Sch\"offler}}, \bibinfo {author} {\bibfnamefont {Xinhua}\
  \bibnamefont {Xie}}, \bibinfo {author} {\bibfnamefont {Philipp}\ \bibnamefont
  {Wustelt}}, \bibinfo {author} {\bibfnamefont {Max}\ \bibnamefont {M\"oller}},
  \bibinfo {author} {\bibfnamefont {Stefan}\ \bibnamefont {Roither}}, \bibinfo
  {author} {\bibfnamefont {Daniil}\ \bibnamefont {Kartashov}}, \bibinfo
  {author} {\bibfnamefont {A.~Max}\ \bibnamefont {Sayler}}, \bibinfo {author}
  {\bibfnamefont {Andrius}\ \bibnamefont {Baltuska}}, \bibinfo {author}
  {\bibfnamefont {Gerhard~G.}\ \bibnamefont {Paulus}}, \ and\ \bibinfo {author}
  {\bibfnamefont {Markus}\ \bibnamefont {Kitzler}},\ }\bibfield  {title}
  {\enquote {\bibinfo {title} {Laser-subcycle control of sequential
  double-ionization dynamics of helium},}\ }\href {\doibase
  10.1103/PhysRevA.93.063421} {\bibfield  {journal} {\bibinfo  {journal}
  {Physical Review A}\ }\textbf {\bibinfo {volume} {93}},\ \bibinfo {pages}
  {063421} (\bibinfo {year} {2016})}\BibitemShut {NoStop}%
\bibitem [{\citenamefont {{Bergues}}\ \emph {et~al.}(2015)\citenamefont
  {{Bergues}}, \citenamefont {{Kübel}}, \citenamefont {{Kling}}, \citenamefont
  {{Burger}},\ and\ \citenamefont {{Kling}}}]{Bergues:2015}%
  \BibitemOpen
  \bibfield  {author} {\bibinfo {author} {\bibfnamefont {B.}~\bibnamefont
  {{Bergues}}}, \bibinfo {author} {\bibfnamefont {M.}~\bibnamefont {{Kübel}}},
  \bibinfo {author} {\bibfnamefont {N.~G.}\ \bibnamefont {{Kling}}}, \bibinfo
  {author} {\bibfnamefont {C.}~\bibnamefont {{Burger}}}, \ and\ \bibinfo
  {author} {\bibfnamefont {M.~F.}\ \bibnamefont {{Kling}}},\ }\bibfield
  {title} {\enquote {\bibinfo {title} {Single-cycle non-sequential double
  ionization},}\ }\href {\doibase 10.1109/JSTQE.2015.2443976} {\bibfield
  {journal} {\bibinfo  {journal} {IEEE Journal of Selected Topics in Quantum
  Electronics}\ }\textbf {\bibinfo {volume} {21}},\ \bibinfo {pages} {1--9}
  (\bibinfo {year} {2015})}\BibitemShut {NoStop}%
\bibitem [{\citenamefont {K\"ubel}\ \emph
  {et~al.}(2016{\natexlab{a}})\citenamefont {K\"ubel}, \citenamefont {Burger},
  \citenamefont {Kling}, \citenamefont {Pischke}, \citenamefont {Beaufore},
  \citenamefont {Ben-Itzhak}, \citenamefont {Paulus}, \citenamefont {Ullrich},
  \citenamefont {Pfeifer}, \citenamefont {Moshammer}, \citenamefont {Kling},\
  and\ \citenamefont {Bergues}}]{Kubel:2016}%
  \BibitemOpen
  \bibfield  {author} {\bibinfo {author} {\bibfnamefont {M.}~\bibnamefont
  {K\"ubel}}, \bibinfo {author} {\bibfnamefont {C.}~\bibnamefont {Burger}},
  \bibinfo {author} {\bibfnamefont {Nora~G.}\ \bibnamefont {Kling}}, \bibinfo
  {author} {\bibfnamefont {T.}~\bibnamefont {Pischke}}, \bibinfo {author}
  {\bibfnamefont {L.}~\bibnamefont {Beaufore}}, \bibinfo {author}
  {\bibfnamefont {I.}~\bibnamefont {Ben-Itzhak}}, \bibinfo {author}
  {\bibfnamefont {G.~G.}\ \bibnamefont {Paulus}}, \bibinfo {author}
  {\bibfnamefont {J.}~\bibnamefont {Ullrich}}, \bibinfo {author} {\bibfnamefont
  {T.}~\bibnamefont {Pfeifer}}, \bibinfo {author} {\bibfnamefont
  {R.}~\bibnamefont {Moshammer}}, \bibinfo {author} {\bibfnamefont {M.~F.}\
  \bibnamefont {Kling}}, \ and\ \bibinfo {author} {\bibfnamefont
  {B.}~\bibnamefont {Bergues}},\ }\bibfield  {title} {\enquote {\bibinfo
  {title} {Complete characterization of single-cycle double ionization of argon
  from the nonsequential to the sequential ionization regime},}\ }\href
  {\doibase 10.1103/PhysRevA.93.053422} {\bibfield  {journal} {\bibinfo
  {journal} {Physical Review A}\ }\textbf {\bibinfo {volume} {93}},\ \bibinfo
  {pages} {053422} (\bibinfo {year} {2016}{\natexlab{a}})}\BibitemShut
  {NoStop}%
\bibitem [{\citenamefont {Alnaser}\ \emph {et~al.}(2014)\citenamefont
  {Alnaser}, \citenamefont {K{\"u}bel}, \citenamefont {Siemering},
  \citenamefont {Bergues}, \citenamefont {Kling}, \citenamefont {Betsch},
  \citenamefont {Deng}, \citenamefont {Schmidt}, \citenamefont {Alahmed},
  \citenamefont {Azzeer}, \citenamefont {Ullrich}, \citenamefont {Ben-Itzhak},
  \citenamefont {Moshammer}, \citenamefont {Kleineberg}, \citenamefont
  {Krausz}, \citenamefont {de~Vivie-Riedle},\ and\ \citenamefont
  {Kling}}]{Alnaser:2014}%
  \BibitemOpen
  \bibfield  {author} {\bibinfo {author} {\bibfnamefont {A.~S.}\ \bibnamefont
  {Alnaser}}, \bibinfo {author} {\bibfnamefont {M.}~\bibnamefont {K{\"u}bel}},
  \bibinfo {author} {\bibfnamefont {R.}~\bibnamefont {Siemering}}, \bibinfo
  {author} {\bibfnamefont {B.}~\bibnamefont {Bergues}}, \bibinfo {author}
  {\bibfnamefont {Nora~G.}\ \bibnamefont {Kling}}, \bibinfo {author}
  {\bibfnamefont {K.~J.}\ \bibnamefont {Betsch}}, \bibinfo {author}
  {\bibfnamefont {Y.}~\bibnamefont {Deng}}, \bibinfo {author} {\bibfnamefont
  {J.}~\bibnamefont {Schmidt}}, \bibinfo {author} {\bibfnamefont {Z.~A.}\
  \bibnamefont {Alahmed}}, \bibinfo {author} {\bibfnamefont {A.~M.}\
  \bibnamefont {Azzeer}}, \bibinfo {author} {\bibfnamefont {J.}~\bibnamefont
  {Ullrich}}, \bibinfo {author} {\bibfnamefont {I.}~\bibnamefont {Ben-Itzhak}},
  \bibinfo {author} {\bibfnamefont {R.}~\bibnamefont {Moshammer}}, \bibinfo
  {author} {\bibfnamefont {U.}~\bibnamefont {Kleineberg}}, \bibinfo {author}
  {\bibfnamefont {F.}~\bibnamefont {Krausz}}, \bibinfo {author} {\bibfnamefont
  {R.}~\bibnamefont {de~Vivie-Riedle}}, \ and\ \bibinfo {author} {\bibfnamefont
  {M.~F.}\ \bibnamefont {Kling}},\ }\bibfield  {title} {\enquote {\bibinfo
  {title} {Subfemtosecond steering of hydrocarbon deprotonation through
  superposition of vibrational modes},}\ }\href
  {https://doi.org/10.1038/ncomms4800} {\bibfield  {journal} {\bibinfo
  {journal} {Nature Communications}\ }\textbf {\bibinfo {volume} {5}},\
  \bibinfo {pages} {3800--3805} (\bibinfo {year} {2014})}\BibitemShut {NoStop}%
\bibitem [{\citenamefont {Miura}\ \emph {et~al.}(2014)\citenamefont {Miura},
  \citenamefont {Ando}, \citenamefont {Ootaka}, \citenamefont {Iwasaki},
  \citenamefont {Xu}, \citenamefont {Okino}, \citenamefont {Yamanouchi},
  \citenamefont {Hoff}, \citenamefont {Rathje}, \citenamefont {Paulus},
  \citenamefont {Kitzler}, \citenamefont {Baltu{\v s}ka}, \citenamefont
  {Sansone},\ and\ \citenamefont {Nisoli}}]{Miura:2014}%
  \BibitemOpen
  \bibfield  {author} {\bibinfo {author} {\bibfnamefont {Shun}\ \bibnamefont
  {Miura}}, \bibinfo {author} {\bibfnamefont {Toshiaki}\ \bibnamefont {Ando}},
  \bibinfo {author} {\bibfnamefont {Kazuki}\ \bibnamefont {Ootaka}}, \bibinfo
  {author} {\bibfnamefont {Atsushi}\ \bibnamefont {Iwasaki}}, \bibinfo {author}
  {\bibfnamefont {Huailiang}\ \bibnamefont {Xu}}, \bibinfo {author}
  {\bibfnamefont {Tomoya}\ \bibnamefont {Okino}}, \bibinfo {author}
  {\bibfnamefont {Kaoru}\ \bibnamefont {Yamanouchi}}, \bibinfo {author}
  {\bibfnamefont {Dominik}\ \bibnamefont {Hoff}}, \bibinfo {author}
  {\bibfnamefont {Tim}\ \bibnamefont {Rathje}}, \bibinfo {author}
  {\bibfnamefont {Gerhard~G.}\ \bibnamefont {Paulus}}, \bibinfo {author}
  {\bibfnamefont {Markus}\ \bibnamefont {Kitzler}}, \bibinfo {author}
  {\bibfnamefont {Andrius}\ \bibnamefont {Baltu{\v s}ka}}, \bibinfo {author}
  {\bibfnamefont {Giuseppe}\ \bibnamefont {Sansone}}, \ and\ \bibinfo {author}
  {\bibfnamefont {Mauro}\ \bibnamefont {Nisoli}},\ }\bibfield  {title}
  {\enquote {\bibinfo {title} {Carrier-envelope-phase dependence of asymmetric
  cd bond breaking in c2d2 in an intense few-cycle laser field},}\ }\href
  {\doibase 10.1016/j.cplett.2014.01.045} {\bibfield  {journal} {\bibinfo
  {journal} {Chemical Physics Letters}\ }\textbf {\bibinfo {volume}
  {595-596}},\ \bibinfo {pages} {61 -- 66} (\bibinfo {year}
  {2014})}\BibitemShut {NoStop}%
\bibitem [{\citenamefont {K\"ubel}\ \emph
  {et~al.}(2016{\natexlab{b}})\citenamefont {K\"ubel}, \citenamefont
  {Siemering}, \citenamefont {Burger}, \citenamefont {Kling}, \citenamefont
  {Li}, \citenamefont {Alnaser}, \citenamefont {Bergues}, \citenamefont
  {Zherebtsov}, \citenamefont {Azzeer}, \citenamefont {Ben-Itzhak},
  \citenamefont {Moshammer}, \citenamefont {de~Vivie-Riedle},\ and\
  \citenamefont {Kling}}]{Kubel:2016-2}%
  \BibitemOpen
  \bibfield  {author} {\bibinfo {author} {\bibfnamefont {M.}~\bibnamefont
  {K\"ubel}}, \bibinfo {author} {\bibfnamefont {R.}~\bibnamefont {Siemering}},
  \bibinfo {author} {\bibfnamefont {C.}~\bibnamefont {Burger}}, \bibinfo
  {author} {\bibfnamefont {Nora~G.}\ \bibnamefont {Kling}}, \bibinfo {author}
  {\bibfnamefont {H.}~\bibnamefont {Li}}, \bibinfo {author} {\bibfnamefont
  {A.~S.}\ \bibnamefont {Alnaser}}, \bibinfo {author} {\bibfnamefont
  {B.}~\bibnamefont {Bergues}}, \bibinfo {author} {\bibfnamefont
  {S.}~\bibnamefont {Zherebtsov}}, \bibinfo {author} {\bibfnamefont {A.~M.}\
  \bibnamefont {Azzeer}}, \bibinfo {author} {\bibfnamefont {I.}~\bibnamefont
  {Ben-Itzhak}}, \bibinfo {author} {\bibfnamefont {R.}~\bibnamefont
  {Moshammer}}, \bibinfo {author} {\bibfnamefont {R.}~\bibnamefont
  {de~Vivie-Riedle}}, \ and\ \bibinfo {author} {\bibfnamefont {M.~F.}\
  \bibnamefont {Kling}},\ }\bibfield  {title} {\enquote {\bibinfo {title}
  {Steering proton migration in hydrocarbons using intense few-cycle laser
  fields},}\ }\href {\doibase 10.1103/PhysRevLett.116.193001} {\bibfield
  {journal} {\bibinfo  {journal} {Physical Review Letters}\ }\textbf {\bibinfo
  {volume} {116}},\ \bibinfo {pages} {193001} (\bibinfo {year}
  {2016}{\natexlab{b}})}\BibitemShut {NoStop}%
\bibitem [{\citenamefont {Süßmann}\ \emph {et~al.}(2011)\citenamefont
  {Süßmann}, \citenamefont {Zherebtsov}, \citenamefont {Plenge},
  \citenamefont {Johnson}, \citenamefont {Kübel}, \citenamefont {Sayler},
  \citenamefont {Mondes}, \citenamefont {Graf}, \citenamefont {Rühl},
  \citenamefont {Paulus}, \citenamefont {Schmischke}, \citenamefont
  {Swrschek},\ and\ \citenamefont {Kling}}]{Sussmann:2011}%
  \BibitemOpen
  \bibfield  {author} {\bibinfo {author} {\bibfnamefont {F.}~\bibnamefont
  {Süßmann}}, \bibinfo {author} {\bibfnamefont {S.}~\bibnamefont
  {Zherebtsov}}, \bibinfo {author} {\bibfnamefont {J.}~\bibnamefont {Plenge}},
  \bibinfo {author} {\bibfnamefont {Nora~G.}\ \bibnamefont {Johnson}}, \bibinfo
  {author} {\bibfnamefont {M.}~\bibnamefont {Kübel}}, \bibinfo {author}
  {\bibfnamefont {A.~M.}\ \bibnamefont {Sayler}}, \bibinfo {author}
  {\bibfnamefont {V.}~\bibnamefont {Mondes}}, \bibinfo {author} {\bibfnamefont
  {C.}~\bibnamefont {Graf}}, \bibinfo {author} {\bibfnamefont {E.}~\bibnamefont
  {Rühl}}, \bibinfo {author} {\bibfnamefont {G.~G.}\ \bibnamefont {Paulus}},
  \bibinfo {author} {\bibfnamefont {D.}~\bibnamefont {Schmischke}}, \bibinfo
  {author} {\bibfnamefont {P.}~\bibnamefont {Swrschek}}, \ and\ \bibinfo
  {author} {\bibfnamefont {M.~F.}\ \bibnamefont {Kling}},\ }\bibfield  {title}
  {\enquote {\bibinfo {title} {Single-shot velocity-map imaging of attosecond
  light-field control at kilohertz rate},}\ }\href {\doibase 10.1063/1.3639333}
  {\bibfield  {journal} {\bibinfo  {journal} {Review of Scientific
  Instruments}\ }\textbf {\bibinfo {volume} {82}},\ \bibinfo {pages} {093109}
  (\bibinfo {year} {2011})},\ \Eprint
  {http://arxiv.org/abs/https://doi.org/10.1063/1.3639333}
  {https://doi.org/10.1063/1.3639333} \BibitemShut {NoStop}%
\bibitem [{\citenamefont {S{\"u}{\ss}mann}\ \emph {et~al.}(2015)\citenamefont
  {S{\"u}{\ss}mann}, \citenamefont {Seiffert}, \citenamefont {Zherebtsov},
  \citenamefont {Mondes}, \citenamefont {Stierle}, \citenamefont {Arbeiter},
  \citenamefont {Plenge}, \citenamefont {Rupp}, \citenamefont {Peltz},
  \citenamefont {Kessel}, \citenamefont {Trushin}, \citenamefont {Ahn},
  \citenamefont {Kim}, \citenamefont {Graf}, \citenamefont {R{\"u}hl},
  \citenamefont {Kling},\ and\ \citenamefont {Fennel}}]{Sussmann:2015}%
  \BibitemOpen
  \bibfield  {author} {\bibinfo {author} {\bibfnamefont {F.}~\bibnamefont
  {S{\"u}{\ss}mann}}, \bibinfo {author} {\bibfnamefont {L.}~\bibnamefont
  {Seiffert}}, \bibinfo {author} {\bibfnamefont {S.}~\bibnamefont
  {Zherebtsov}}, \bibinfo {author} {\bibfnamefont {V.}~\bibnamefont {Mondes}},
  \bibinfo {author} {\bibfnamefont {J.}~\bibnamefont {Stierle}}, \bibinfo
  {author} {\bibfnamefont {M.}~\bibnamefont {Arbeiter}}, \bibinfo {author}
  {\bibfnamefont {J.}~\bibnamefont {Plenge}}, \bibinfo {author} {\bibfnamefont
  {P.}~\bibnamefont {Rupp}}, \bibinfo {author} {\bibfnamefont {C.}~\bibnamefont
  {Peltz}}, \bibinfo {author} {\bibfnamefont {A.}~\bibnamefont {Kessel}},
  \bibinfo {author} {\bibfnamefont {S.~A.}\ \bibnamefont {Trushin}}, \bibinfo
  {author} {\bibfnamefont {B.}~\bibnamefont {Ahn}}, \bibinfo {author}
  {\bibfnamefont {D.}~\bibnamefont {Kim}}, \bibinfo {author} {\bibfnamefont
  {C.}~\bibnamefont {Graf}}, \bibinfo {author} {\bibfnamefont {E.}~\bibnamefont
  {R{\"u}hl}}, \bibinfo {author} {\bibfnamefont {M.~F.}\ \bibnamefont {Kling}},
  \ and\ \bibinfo {author} {\bibfnamefont {T.}~\bibnamefont {Fennel}},\
  }\bibfield  {title} {\enquote {\bibinfo {title} {Field propagation-induced
  directionality of carrier-envelope phase-controlled photoemission from
  nanospheres},}\ }\href {https://doi.org/10.1038/ncomms8944} {\bibfield
  {journal} {\bibinfo  {journal} {Nature Communications}\ }\textbf {\bibinfo
  {volume} {6}},\ \bibinfo {pages} {7944--7952} (\bibinfo {year}
  {2015})}\BibitemShut {NoStop}%
\bibitem [{\citenamefont {Kormin}\ \emph {et~al.}(2018)\citenamefont {Kormin},
  \citenamefont {Borot}, \citenamefont {Ma}, \citenamefont {Dallari},
  \citenamefont {Bergues}, \citenamefont {Aladi}, \citenamefont {F{\"o}ldes},\
  and\ \citenamefont {Veisz}}]{Kormin:2018}%
  \BibitemOpen
  \bibfield  {author} {\bibinfo {author} {\bibfnamefont {Dmitrii}\ \bibnamefont
  {Kormin}}, \bibinfo {author} {\bibfnamefont {Antonin}\ \bibnamefont {Borot}},
  \bibinfo {author} {\bibfnamefont {Guangjin}\ \bibnamefont {Ma}}, \bibinfo
  {author} {\bibfnamefont {William}\ \bibnamefont {Dallari}}, \bibinfo {author}
  {\bibfnamefont {Boris}\ \bibnamefont {Bergues}}, \bibinfo {author}
  {\bibfnamefont {M{\'a}rk}\ \bibnamefont {Aladi}}, \bibinfo {author}
  {\bibfnamefont {Istv{\'a}n~B.}\ \bibnamefont {F{\"o}ldes}}, \ and\ \bibinfo
  {author} {\bibfnamefont {Laszlo}\ \bibnamefont {Veisz}},\ }\bibfield  {title}
  {\enquote {\bibinfo {title} {Spectral interferometry with waveform-dependent
  relativistic high-order harmonics from plasma surfaces},}\ }\href {\doibase
  10.1038/s41467-018-07421-5} {\bibfield  {journal} {\bibinfo  {journal}
  {Nature Communications}\ }\textbf {\bibinfo {volume} {9}},\ \bibinfo {pages}
  {4992} (\bibinfo {year} {2018})}\BibitemShut {NoStop}%
\bibitem [{\citenamefont {Kessel}\ \emph {et~al.}(2018)\citenamefont {Kessel},
  \citenamefont {Leshchenko}, \citenamefont {Jahn}, \citenamefont {Kr\"{u}ger},
  \citenamefont {M\"{u}nzer}, \citenamefont {Schwarz}, \citenamefont {Pervak},
  \citenamefont {Trubetskov}, \citenamefont {Trushin}, \citenamefont {Krausz},
  \citenamefont {Major},\ and\ \citenamefont {Karsch}}]{Kessel:2018}%
  \BibitemOpen
  \bibfield  {author} {\bibinfo {author} {\bibfnamefont {Alexander}\
  \bibnamefont {Kessel}}, \bibinfo {author} {\bibfnamefont {Vyacheslav~E.}\
  \bibnamefont {Leshchenko}}, \bibinfo {author} {\bibfnamefont {Olga}\
  \bibnamefont {Jahn}}, \bibinfo {author} {\bibfnamefont {Mathias}\
  \bibnamefont {Kr\"{u}ger}}, \bibinfo {author} {\bibfnamefont {Andreas}\
  \bibnamefont {M\"{u}nzer}}, \bibinfo {author} {\bibfnamefont {Alexander}\
  \bibnamefont {Schwarz}}, \bibinfo {author} {\bibfnamefont {Vladimir}\
  \bibnamefont {Pervak}}, \bibinfo {author} {\bibfnamefont {Michael}\
  \bibnamefont {Trubetskov}}, \bibinfo {author} {\bibfnamefont {Sergei~A.}\
  \bibnamefont {Trushin}}, \bibinfo {author} {\bibfnamefont {Ferenc}\
  \bibnamefont {Krausz}}, \bibinfo {author} {\bibfnamefont {Zsuzsanna}\
  \bibnamefont {Major}}, \ and\ \bibinfo {author} {\bibfnamefont {Stefan}\
  \bibnamefont {Karsch}},\ }\bibfield  {title} {\enquote {\bibinfo {title}
  {Relativistic few-cycle pulses with high contrast from picosecond-pumped
  opcpa},}\ }\href {\doibase 10.1364/OPTICA.5.000434} {\bibfield  {journal}
  {\bibinfo  {journal} {Optica}\ }\textbf {\bibinfo {volume} {5}},\ \bibinfo
  {pages} {434--442} (\bibinfo {year} {2018})}\BibitemShut {NoStop}%
\bibitem [{\citenamefont {Tate}\ \emph {et~al.}(2007)\citenamefont {Tate},
  \citenamefont {Auguste}, \citenamefont {Muller}, \citenamefont {Sali\`eres},
  \citenamefont {Agostini},\ and\ \citenamefont {DiMauro}}]{Tate:2007}%
  \BibitemOpen
  \bibfield  {author} {\bibinfo {author} {\bibfnamefont {J.}~\bibnamefont
  {Tate}}, \bibinfo {author} {\bibfnamefont {T.}~\bibnamefont {Auguste}},
  \bibinfo {author} {\bibfnamefont {H.~G.}\ \bibnamefont {Muller}}, \bibinfo
  {author} {\bibfnamefont {P.}~\bibnamefont {Sali\`eres}}, \bibinfo {author}
  {\bibfnamefont {P.}~\bibnamefont {Agostini}}, \ and\ \bibinfo {author}
  {\bibfnamefont {L.~F.}\ \bibnamefont {DiMauro}},\ }\bibfield  {title}
  {\enquote {\bibinfo {title} {Scaling of wave-packet dynamics in an intense
  midinfrared field},}\ }\href {\doibase 10.1103/PhysRevLett.98.013901}
  {\bibfield  {journal} {\bibinfo  {journal} {Physical Review Letters}\
  }\textbf {\bibinfo {volume} {98}},\ \bibinfo {pages} {013901} (\bibinfo
  {year} {2007})}\BibitemShut {NoStop}%
\bibitem [{\citenamefont {Schiffrin}\ \emph {et~al.}(2012)\citenamefont
  {Schiffrin}, \citenamefont {Paasch-Colberg}, \citenamefont {Karpowicz},
  \citenamefont {Apalkov}, \citenamefont {Gerster}, \citenamefont
  {M{\"u}hlbrandt}, \citenamefont {Korbman}, \citenamefont {Reichert},
  \citenamefont {Schultze}, \citenamefont {Holzner}, \citenamefont {Barth},
  \citenamefont {Kienberger}, \citenamefont {Ernstorfer}, \citenamefont
  {Yakovlev}, \citenamefont {Stockman},\ and\ \citenamefont
  {Krausz}}]{Schiffrin:2013}%
  \BibitemOpen
  \bibfield  {author} {\bibinfo {author} {\bibfnamefont {Agustin}\ \bibnamefont
  {Schiffrin}}, \bibinfo {author} {\bibfnamefont {Tim}\ \bibnamefont
  {Paasch-Colberg}}, \bibinfo {author} {\bibfnamefont {Nicholas}\ \bibnamefont
  {Karpowicz}}, \bibinfo {author} {\bibfnamefont {Vadym}\ \bibnamefont
  {Apalkov}}, \bibinfo {author} {\bibfnamefont {Daniel}\ \bibnamefont
  {Gerster}}, \bibinfo {author} {\bibfnamefont {Sascha}\ \bibnamefont
  {M{\"u}hlbrandt}}, \bibinfo {author} {\bibfnamefont {Michael}\ \bibnamefont
  {Korbman}}, \bibinfo {author} {\bibfnamefont {Joachim}\ \bibnamefont
  {Reichert}}, \bibinfo {author} {\bibfnamefont {Martin}\ \bibnamefont
  {Schultze}}, \bibinfo {author} {\bibfnamefont {Simon}\ \bibnamefont
  {Holzner}}, \bibinfo {author} {\bibfnamefont {Johannes~V.}\ \bibnamefont
  {Barth}}, \bibinfo {author} {\bibfnamefont {Reinhard}\ \bibnamefont
  {Kienberger}}, \bibinfo {author} {\bibfnamefont {Ralph}\ \bibnamefont
  {Ernstorfer}}, \bibinfo {author} {\bibfnamefont {Vladislav~S.}\ \bibnamefont
  {Yakovlev}}, \bibinfo {author} {\bibfnamefont {Mark~I.}\ \bibnamefont
  {Stockman}}, \ and\ \bibinfo {author} {\bibfnamefont {Ferenc}\ \bibnamefont
  {Krausz}},\ }\bibfield  {title} {\enquote {\bibinfo {title}
  {Optical-field-induced current in dielectrics},}\ }\href
  {https://doi.org/10.1038/nature11567} {\bibfield  {journal} {\bibinfo
  {journal} {Nature}\ }\textbf {\bibinfo {volume} {493}},\ \bibinfo {pages}
  {70--74} (\bibinfo {year} {2012})}\BibitemShut {NoStop}%
\bibitem [{\citenamefont {Paasch-Colberg}\ \emph {et~al.}(2014)\citenamefont
  {Paasch-Colberg}, \citenamefont {Schiffrin}, \citenamefont {Karpowicz},
  \citenamefont {Kruchinin}, \citenamefont {Saglam}, \citenamefont {Keiber},
  \citenamefont {Razskazovskaya}, \citenamefont {M{\"u}hlbrandt}, \citenamefont
  {Alnaser}, \citenamefont {K{\"u}bel}, \citenamefont {Apalkov}, \citenamefont
  {Gerster}, \citenamefont {Reichert}, \citenamefont {Wittmann}, \citenamefont
  {Barth}, \citenamefont {Stockman}, \citenamefont {Ernstorfer}, \citenamefont
  {Yakovlev}, \citenamefont {Kienberger},\ and\ \citenamefont
  {Krausz}}]{Paasch-Colberg:2014}%
  \BibitemOpen
  \bibfield  {author} {\bibinfo {author} {\bibfnamefont {Tim}\ \bibnamefont
  {Paasch-Colberg}}, \bibinfo {author} {\bibfnamefont {Agustin}\ \bibnamefont
  {Schiffrin}}, \bibinfo {author} {\bibfnamefont {Nicholas}\ \bibnamefont
  {Karpowicz}}, \bibinfo {author} {\bibfnamefont {Stanislav}\ \bibnamefont
  {Kruchinin}}, \bibinfo {author} {\bibfnamefont {{\"O}zge}\ \bibnamefont
  {Saglam}}, \bibinfo {author} {\bibfnamefont {Sabine}\ \bibnamefont {Keiber}},
  \bibinfo {author} {\bibfnamefont {Olga}\ \bibnamefont {Razskazovskaya}},
  \bibinfo {author} {\bibfnamefont {Sascha}\ \bibnamefont {M{\"u}hlbrandt}},
  \bibinfo {author} {\bibfnamefont {Ali}\ \bibnamefont {Alnaser}}, \bibinfo
  {author} {\bibfnamefont {Matthias}\ \bibnamefont {K{\"u}bel}}, \bibinfo
  {author} {\bibfnamefont {Vadym}\ \bibnamefont {Apalkov}}, \bibinfo {author}
  {\bibfnamefont {Daniel}\ \bibnamefont {Gerster}}, \bibinfo {author}
  {\bibfnamefont {Joachim}\ \bibnamefont {Reichert}}, \bibinfo {author}
  {\bibfnamefont {Tibor}\ \bibnamefont {Wittmann}}, \bibinfo {author}
  {\bibfnamefont {Johannes~V.}\ \bibnamefont {Barth}}, \bibinfo {author}
  {\bibfnamefont {Mark~I.}\ \bibnamefont {Stockman}}, \bibinfo {author}
  {\bibfnamefont {Ralph}\ \bibnamefont {Ernstorfer}}, \bibinfo {author}
  {\bibfnamefont {Vladislav~S.}\ \bibnamefont {Yakovlev}}, \bibinfo {author}
  {\bibfnamefont {Reinhard}\ \bibnamefont {Kienberger}}, \ and\ \bibinfo
  {author} {\bibfnamefont {Ferenc}\ \bibnamefont {Krausz}},\ }\bibfield
  {title} {\enquote {\bibinfo {title} {Solid-state light-phase detector},}\
  }\href {https://doi.org/10.1038/nphoton.2013.348} {\bibfield  {journal}
  {\bibinfo  {journal} {Nature Photonics}\ }\textbf {\bibinfo {volume} {8}},\
  \bibinfo {pages} {214--218} (\bibinfo {year} {2014})}\BibitemShut {NoStop}%
\bibitem [{\citenamefont {Bergues}(2012)}]{Bergues:2012-2}%
  \BibitemOpen
  \bibfield  {author} {\bibinfo {author} {\bibfnamefont {Boris}\ \bibnamefont
  {Bergues}},\ }\bibfield  {title} {\enquote {\bibinfo {title} {The
  circular-polarization phase-meter},}\ }\href {\doibase 10.1364/OE.20.025317}
  {\bibfield  {journal} {\bibinfo  {journal} {Optics Express}\ }\textbf
  {\bibinfo {volume} {20}},\ \bibinfo {pages} {25317--25324} (\bibinfo {year}
  {2012})}\BibitemShut {NoStop}%
\bibitem [{\citenamefont {Fukahori}\ \emph {et~al.}(2017)\citenamefont
  {Fukahori}, \citenamefont {Ando}, \citenamefont {Miura}, \citenamefont
  {Kanya}, \citenamefont {Yamanouchi}, \citenamefont {Rathje},\ and\
  \citenamefont {Paulus}}]{Fukahori:2017}%
  \BibitemOpen
  \bibfield  {author} {\bibinfo {author} {\bibfnamefont {Shinichi}\
  \bibnamefont {Fukahori}}, \bibinfo {author} {\bibfnamefont {Toshiaki}\
  \bibnamefont {Ando}}, \bibinfo {author} {\bibfnamefont {Shun}\ \bibnamefont
  {Miura}}, \bibinfo {author} {\bibfnamefont {Reika}\ \bibnamefont {Kanya}},
  \bibinfo {author} {\bibfnamefont {Kaoru}\ \bibnamefont {Yamanouchi}},
  \bibinfo {author} {\bibfnamefont {Tim}\ \bibnamefont {Rathje}}, \ and\
  \bibinfo {author} {\bibfnamefont {Gerhard~G.}\ \bibnamefont {Paulus}},\
  }\bibfield  {title} {\enquote {\bibinfo {title} {Determination of the
  absolute carrier-envelope phase by angle-resolved photoelectron spectra of ar
  by intense circularly polarized few-cycle pulses},}\ }\href {\doibase
  10.1103/PhysRevA.95.053410} {\bibfield  {journal} {\bibinfo  {journal}
  {Physical Review A}\ }\textbf {\bibinfo {volume} {95}},\ \bibinfo {pages}
  {053410} (\bibinfo {year} {2017})}\BibitemShut {NoStop}%
\bibitem [{\citenamefont {Debrah}\ \emph {et~al.}(2019)\citenamefont {Debrah},
  \citenamefont {Stewart}, \citenamefont {Basnayake}, \citenamefont {Tisch},
  \citenamefont {Lee},\ and\ \citenamefont {Li}}]{Debrah:2019}%
  \BibitemOpen
  \bibfield  {author} {\bibinfo {author} {\bibfnamefont {Duke~A.}\ \bibnamefont
  {Debrah}}, \bibinfo {author} {\bibfnamefont {Gabriel~A.}\ \bibnamefont
  {Stewart}}, \bibinfo {author} {\bibfnamefont {Gihan}\ \bibnamefont
  {Basnayake}}, \bibinfo {author} {\bibfnamefont {John W.~G.}\ \bibnamefont
  {Tisch}}, \bibinfo {author} {\bibfnamefont {Suk~Kyoung}\ \bibnamefont {Lee}},
  \ and\ \bibinfo {author} {\bibfnamefont {Wen}\ \bibnamefont {Li}},\
  }\bibfield  {title} {\enquote {\bibinfo {title} {Direct in-situ single-shot
  measurements of the absolute carrier-envelope phases of ultrashort pulses},}\
  }\href {\doibase 10.1364/OL.44.003582} {\bibfield  {journal} {\bibinfo
  {journal} {Opt. Lett.}\ }\textbf {\bibinfo {volume} {44}},\ \bibinfo {pages}
  {3582--3585} (\bibinfo {year} {2019})}\BibitemShut {NoStop}%
\bibitem [{\citenamefont {Kre{\ss}}\ \emph {et~al.}(2006)\citenamefont
  {Kre{\ss}}, \citenamefont {L{\"o}ffler}, \citenamefont {Thomson},
  \citenamefont {D{\"o}rner}, \citenamefont {Gimpel}, \citenamefont {Zrost},
  \citenamefont {Ergler}, \citenamefont {Moshammer}, \citenamefont {Morgner},
  \citenamefont {Ullrich},\ and\ \citenamefont {Roskos}}]{Kress:2006}%
  \BibitemOpen
  \bibfield  {author} {\bibinfo {author} {\bibfnamefont {Markus}\ \bibnamefont
  {Kre{\ss}}}, \bibinfo {author} {\bibfnamefont {Torsten}\ \bibnamefont
  {L{\"o}ffler}}, \bibinfo {author} {\bibfnamefont {Mark~D.}\ \bibnamefont
  {Thomson}}, \bibinfo {author} {\bibfnamefont {Reinhard}\ \bibnamefont
  {D{\"o}rner}}, \bibinfo {author} {\bibfnamefont {Hartmut}\ \bibnamefont
  {Gimpel}}, \bibinfo {author} {\bibfnamefont {Karl}\ \bibnamefont {Zrost}},
  \bibinfo {author} {\bibfnamefont {Thorsten}\ \bibnamefont {Ergler}}, \bibinfo
  {author} {\bibfnamefont {Robert}\ \bibnamefont {Moshammer}}, \bibinfo
  {author} {\bibfnamefont {Uwe}\ \bibnamefont {Morgner}}, \bibinfo {author}
  {\bibfnamefont {Joachim}\ \bibnamefont {Ullrich}}, \ and\ \bibinfo {author}
  {\bibfnamefont {Hartmut~G.}\ \bibnamefont {Roskos}},\ }\bibfield  {title}
  {\enquote {\bibinfo {title} {Determination of the carrier-envelope phase of
  few-cycle laser pulses with terahertz-emission spectroscopy},}\ }\href
  {\doibase 10.1038/nphys286} {\bibfield  {journal} {\bibinfo  {journal}
  {Nature Physics}\ }\textbf {\bibinfo {volume} {2}},\ \bibinfo {pages}
  {327--331} (\bibinfo {year} {2006})}\BibitemShut {NoStop}%
\bibitem [{\citenamefont {Dai}\ \emph {et~al.}(2009)\citenamefont {Dai},
  \citenamefont {Karpowicz},\ and\ \citenamefont {Zhang}}]{Dai:2009}%
  \BibitemOpen
  \bibfield  {author} {\bibinfo {author} {\bibfnamefont {Jianming}\
  \bibnamefont {Dai}}, \bibinfo {author} {\bibfnamefont {Nicholas}\
  \bibnamefont {Karpowicz}}, \ and\ \bibinfo {author} {\bibfnamefont {X.-C.}\
  \bibnamefont {Zhang}},\ }\bibfield  {title} {\enquote {\bibinfo {title}
  {Coherent polarization control of terahertz waves generated from two-color
  laser-induced gas plasma},}\ }\href {\doibase 10.1103/PhysRevLett.103.023001}
  {\bibfield  {journal} {\bibinfo  {journal} {Physical Review Letters}\
  }\textbf {\bibinfo {volume} {103}},\ \bibinfo {pages} {023001} (\bibinfo
  {year} {2009})}\BibitemShut {NoStop}%
\bibitem [{\citenamefont {Bai}\ \emph {et~al.}(2012)\citenamefont {Bai},
  \citenamefont {Li}, \citenamefont {Liu}, \citenamefont {Li}, \citenamefont
  {Xu}, \citenamefont {Yang}, \citenamefont {Zeng}, \citenamefont {Lu},
  \citenamefont {Hu}, \citenamefont {Lei}, \citenamefont {Leng},\ and\
  \citenamefont {Xu}}]{Bai:2012}%
  \BibitemOpen
  \bibfield  {author} {\bibinfo {author} {\bibfnamefont {Y.}~\bibnamefont
  {Bai}}, \bibinfo {author} {\bibfnamefont {C.}~\bibnamefont {Li}}, \bibinfo
  {author} {\bibfnamefont {P.}~\bibnamefont {Liu}}, \bibinfo {author}
  {\bibfnamefont {R.~X.}\ \bibnamefont {Li}}, \bibinfo {author} {\bibfnamefont
  {R.~J.}\ \bibnamefont {Xu}}, \bibinfo {author} {\bibfnamefont
  {H.}~\bibnamefont {Yang}}, \bibinfo {author} {\bibfnamefont {Z.~N.}\
  \bibnamefont {Zeng}}, \bibinfo {author} {\bibfnamefont {H.~H.}\ \bibnamefont
  {Lu}}, \bibinfo {author} {\bibfnamefont {G.~Y.}\ \bibnamefont {Hu}}, \bibinfo
  {author} {\bibfnamefont {A.~L.}\ \bibnamefont {Lei}}, \bibinfo {author}
  {\bibfnamefont {Y.~X.}\ \bibnamefont {Leng}}, \ and\ \bibinfo {author}
  {\bibfnamefont {Z.~Z.}\ \bibnamefont {Xu}},\ }\bibfield  {title} {\enquote
  {\bibinfo {title} {Characterizing the polarity of a few-cycle infrared laser
  pulse},}\ }\href {\doibase 10.1007/s00340-011-4728-4} {\bibfield  {journal}
  {\bibinfo  {journal} {Applied Physics B}\ }\textbf {\bibinfo {volume}
  {106}},\ \bibinfo {pages} {45--49} (\bibinfo {year} {2012})}\BibitemShut
  {NoStop}%
\bibitem [{\citenamefont {K{\" u}bel}\ \emph {et~al.}(2012)\citenamefont {K{\"
  u}bel}, \citenamefont {Betsch}, \citenamefont {Johnson}, \citenamefont
  {Kleineberg}, \citenamefont {Moshammer}, \citenamefont {Ullrich},
  \citenamefont {Paulus}, \citenamefont {Kling},\ and\ \citenamefont
  {Bergues}}]{Kubel:2012}%
  \BibitemOpen
  \bibfield  {author} {\bibinfo {author} {\bibfnamefont {M.}~\bibnamefont {K{\"
  u}bel}}, \bibinfo {author} {\bibfnamefont {K.~J.}\ \bibnamefont {Betsch}},
  \bibinfo {author} {\bibfnamefont {Nora~G.}\ \bibnamefont {Johnson}}, \bibinfo
  {author} {\bibfnamefont {U.}~\bibnamefont {Kleineberg}}, \bibinfo {author}
  {\bibfnamefont {R.}~\bibnamefont {Moshammer}}, \bibinfo {author}
  {\bibfnamefont {J.}~\bibnamefont {Ullrich}}, \bibinfo {author} {\bibfnamefont
  {G.~G.}\ \bibnamefont {Paulus}}, \bibinfo {author} {\bibfnamefont {M.~F.}\
  \bibnamefont {Kling}}, \ and\ \bibinfo {author} {\bibfnamefont
  {B.}~\bibnamefont {Bergues}},\ }\bibfield  {title} {\enquote {\bibinfo
  {title} {Carrier-envelope-phase tagging in measurements with long acquisition
  times},}\ }\href {\doibase 10.1088/1367-2630/14/9/093027} {\bibfield
  {journal} {\bibinfo  {journal} {New Journal of Physics}\ }\textbf {\bibinfo
  {volume} {14}},\ \bibinfo {pages} {093027} (\bibinfo {year}
  {2012})}\BibitemShut {NoStop}%
\end{thebibliography}%

\end{document}